\documentclass[12pt]{article}
\pdfoutput=1 
\usepackage{comment}
\usepackage{graphicx}
\usepackage{bm}
\def\be{\begin{equation}}
\def\ee{\end{equation}}
\def\bea{\begin{eqnarray}}
\def\eea{\end{eqnarray}}
\usepackage{latexsym}
\usepackage{dcolumn}
\usepackage{epsfig,amssymb,euscript}
\usepackage{amsmath}
\usepackage{array,calc,epsfig}
\usepackage{xfrac}

\usepackage[pdfpagelabels]{hyperref}

\usepackage{chngcntr}
\usepackage[noadjust]{cite}

\usepackage{color}

\renewcommand{\to}{\rightarrow}

\def\be{\begin{equation}}
\def\ee{\end{equation}}
\def\ba{\begin{eqnarray}}
\def\ea{\end{eqnarray}}

\def\({\left (}
\def\){\right )}

\let\benn\[
\let\eenn\]

\def\[{\left [}
\def\]{\right ]}

\let\oldlgraf\{ 
\renewcommand{\{}{\left \oldlgraf}

\let\oldrgraf\}
\renewcommand{\}}{\right \oldrgraf}

\counterwithin*{equation}{section}

\begin{document}
\baselineskip=15.5pt
\pagestyle{plain}
\setcounter{page}{1}
\newfont{\namefont}{cmr10}
\newfont{\addfont}{cmti7 scaled 1440}
\newfont{\boldmathfont}{cmbx10}
\newfont{\headfontb}{cmbx10 scaled 1728}
\renewcommand{\theequation}{{\rm\thesection.\arabic{equation}}}
\renewcommand{\thefootnote}{\arabic{footnote}}

\vspace{1cm}

\begin{titlepage}
	\vskip 2cm
	\begin{center}
		{\Large{\bf Holographic Hagedorn Temperature:\\[0.5ex]Strong Coupling, Flavor and $\theta$-angle Effects		}}
	\end{center}
	
	\vskip 10pt
	\begin{center}
		Francesco Bigazzi$^{a}$, Tommaso Canneti$^{a,b}$, Aldo L. Cotrone$^{a,b}$,  Jos\'e Manuel Pen\'in$^{a,b}$
	\end{center}
	\vskip 10pt
	\begin{center}
		\vspace{0.2cm}
		\textit {$^a$ INFN, Sezione di Firenze; Via G. Sansone 1; I-50019 Sesto Fiorentino (Firenze), Italy.
		}\\
		\textit{$^b$ Dipartimento di Fisica e Astronomia, Universit\'a di Firenze; Via G. Sansone 1;\\ I-50019 Sesto Fiorentino (Firenze), Italy.
		}
		\vskip 20pt
		{\small{
				bigazzi@fi.infn.it, canneti@fi.infn.it, cotrone@fi.infn.it, jmanpen@gmail.com}
		}
		
	\end{center}
	
	\vspace{25pt}
	
	\begin{center}
		\textbf{Abstract}
	\end{center}

	\noindent 
	
	We study the Hagedorn temperature $T_H$ of strongly coupled quantum field theories admitting a holographic string or M-theory description in various regimes and scenarios.
	
	In the first part of the paper we propose a ``thermal scalar'' effective approach to the calculation of $T_H$ in eleven-dimensional supergravity.
	The proposal allows to extend the existing results for $T_H$ to the strongly coupled string regime, i.e.~to a previously unexplored regime of field theory parameters where the number of colors $N$ is smaller than (some power of) the 't Hooft coupling $\lambda$.
	We can thus extend the existing results for the $\alpha'$ expansion of the ABJM model, which have a spectacular agreement with predictions from integrability, in a different direction in parameter space.
	In particular, we explicitate the first non-perturbative corrections.
	We also apply the formalism to the Witten-Yang-Mills model, finding that the result for the ratio of $T_H$ with the deconfinement temperature is in the same ballpark of the lattice one for pure Yang-Mills.
	Within the same model, we study the dependence of the Hagedorn temperature on the $\theta$-angle.
	
	In the second part of the paper we analyze the effect of dynamical flavors on $T_H$ in confining theories.
	By studying the few available examples 
	of regular backgrounds dual to confining theories with flavors, we find that generally the effects of flavors is to reduce the value of $T_H$ in units of the square root of the confining string tension. 
	The effect turns out to be milder than the analogous reduction of the critical temperature for deconfinement when the latter is known.

\end{titlepage}

\newpage
\tableofcontents

\section{Introduction}

Many quantum theories have a Hagedorn regime, where the density of states grows exponentially with the energy, causing the divergence of the partition function at some temperature $T_H$ called the Hagedorn temperature.
Examples include both quantum field theories (QFTs), such as Yang-Mills (YM) \cite{Hagedorn:1965st}, as well as string theory \cite{Huang:1970iq,Sundborg:1984uk}.\footnote{For recent progress concerning the density of states of strings in curved spaces see \cite{Canneti:2024iyn}.}

Recently there has been a renewal of interest in this topic, especially in the holographic context \cite{Harmark:2021qma,malda,Urbach:2022xzw,Bigazzi:2022gal,Urbach:2023npi,Bigazzi:2023oqm,Ekhammar:2023glu,Bigazzi:2023hxt,Ekhammar:2023cuj,Harmark:2024ioq,Bigazzi:2024biz}.
In fact, for QFTs admitting a holographic description, the latter provides a convenient framework to study the Hagedorn regime at strong 't Hooft coupling $\lambda \gg 1$ in the planar limit $N \gg 1$.
In recent works, the holographic Hagedorn temperature has been computed either by world-sheet sigma-model methods \cite{Bigazzi:2022gal,Bigazzi:2023oqm}, or by an effective ``thermal scalar'' approach in the Horowitz-Polchinski \cite{Horowitz:1997jc} framework \cite{malda,Urbach:2022xzw,Urbach:2023npi,Ekhammar:2023glu,Ekhammar:2023cuj}, or by a combination of the two \cite{Bigazzi:2023hxt,Harmark:2024ioq,Bigazzi:2024biz}.
In all these cases 
$T_H$ is determined as the temperature at which the lightest string mode winding the compactified temporal direction becomes tachyonic, following Atick and Witten \cite{Atick:1988si}.
Nowadays there exist explicit formulae for $T_H$ for a very general class of backgrounds dual to confining QFTs \cite{Bigazzi:2024biz}. 

These developments have allowed to perform spectacular precision tests of holography, by matching the gravity/string results at NNNLO \cite{Ekhammar:2023glu,Bigazzi:2023hxt,Ekhammar:2023cuj} (i.e.~including third order corrections in $\alpha'$ or equivalently in $1/\lambda$) with Quantum Spectral Curve (QSC) predictions \cite{Harmark:2021qma,Ekhammar:2023cuj} for ${\cal N}=4$ SYM and the ABJM model.\footnote{Let us stress that these tests are at finite temperature, where supersymmetry is broken.} Moreover, they have provided predictions for the NNLO value of $T_H$ in QFTs at strong coupling for which there is no alternative computational framework.

In this work we extend these results in two different directions.
Accordingly, the paper is divided in two parts, which can be read independently.

The existing holographic calculations are all limited to the small dilaton regime, that is $\lambda^a \ll N$ for some $a>0$ in QFT language.
Going beyond this regime requires considering the eleven-dimensional origin of the background.
In M-theory, what in ten dimensions is a (type IIA) string wrapped around the temporal circle becomes a M2 membrane wrapped around the temporal circle and the eleventh dimension (the one related to the dilaton).
In fact, the Hagedorn temperature in flat space can be computed as the temperature at which the lightest mode of the membrane becomes tachyonic \cite{Russo:2001vh}.
This description is in principle 
valid in the limit where the dilaton is large, but it turns out to also reproduce the small dilaton results. 

In the first part of the paper we propose an effective ``thermal scalar'' description of the light membrane mode.
By construction, it coincides with the one for the winding string scalar upon reduction to ten dimensions, reproducing the known results in the literature in the $\lambda^a \ll N$ regime.
Besides, this approach encodes the dependence of $T_H$ on the parameter $\eta \equiv \lambda^a /N$, providing predictions for holographic theories in the large $\eta$ regime.
Moreover, for every fixed value of $\eta$ one can use the same machinery as in ten dimensions, performing an expansion\footnote{The correct expansion parameter depends also on $\eta$. It reduces to $\alpha'$ in the $\eta \ll 1$ limit.} in $\alpha'$ and thus calculating the subleading corrections in $1/\lambda$.
In this way, for theories with a known eleven-dimensional holographic description, one is able to calculate the Hagedorn temperature in a non-trivial region of the two-parameter space $(\lambda, \eta)$.  

This thermal scalar proposal of the membrane mode cannot be tested at the moment even for integrable theories, as the QSC methods are still limited to the small $\eta$ regime.
Needless to say, it would be extremely interesting to have some results from that side.
In particular, we provide a prediction for the value of $T_H$ in the ABJM case \cite{Aharony:2008ug} away from the limit $\eta \rightarrow 0$.
At small $\eta$ the first (analytic) non-perturbative corrections to the known results come into the form of a (D0-brane) instanton gas. 
At large $\eta$ their (analytic) form is power-like as usual in M-theory.
The intermediate region can be analyzed numerically.\footnote{In the intermediate region one could expect, on a general ground,  corrections to our formula, originating both from possible higher curvature corrections to the background and from some approximation performed in \cite{Russo:2001vh}.  
The latter, originating from quartic terms in the membrane theory, should play a role starting at NNLO in the $1/\lambda$ expansion, as it happens in the analogous string calculation in ten dimensions \cite{Bigazzi:2024biz}.}

As a further relevant example, we consider the Witten-YM model (WYM) \cite{Witten:1998zw}.
This is a non-supersymmetric pure YM theory coupled to extra adjoint matter.
We provide the value of $T_H$ for different values of $\eta$ and up to NLO in $1/\lambda$.
In this theory the backreaction of the Yang-Mills $\theta$-angle on the geometry is known \cite{Dubovsky:2011tu, Bigazzi:2015bna}.
Thus, we are also able to study the dependence of the Hagedorn temperature on the $\theta$-angle.

Another interesting consideration in this case comes from comparing the Hagedorn temperature $T_H$ with the critical temperature for deconfinement $T_c$.
At small $\eta$ (when the dual ten-dimensional description is valid), $T_H$ is parametrically larger than $T_c$, $T_H \sim \sqrt{\lambda} T_c$.
But while $T_c$ is constant in $\eta$, $T_H$ is a decreasing function of $\eta$.
As a result, the two temperatures become comparable at a certain value of $\eta$ and $\lambda$.
This regime is particularly interesting since in 
pure Yang-Mills the two temperatures are expected to be quite close \cite{Bringoltz:2005xx}, $T_H/T_c \sim 1.1$.
In the holographic WYM theory, setting $N=3$ and using the rough value of the coupling around $T_c$, our formulae give $T_H/T_c \sim \mathcal O (1)$.

At this point a couple of clarifying comments are in order.
The stringy Hagedorn behavior has been mostly understood in the small dilaton regime ($\eta \ll 1$) where, for example, we will write down the instantonic corrections mentioned above.
At larger values of the dilaton, self-gravitating stringy effects can kick in at smaller temperatures, complicating the analysis \cite{Horowitz:1997jc}.
We have not studied this regime in the stringy setup.
Rather, we have considered the large $\eta$ regime directly in eleven dimensions.
There, the behavior of the wrapped membrane is completely clear.
There appears a tachyonic mode at a certain ``limiting temperature'' (the inverse length of the wrapped circle) set by the Planck length $l_P$.
The situation is precisely as in ten dimensions, where the Hagedorn temperature $T_H$ is calculated, also in flat space, by considering a string wrapped on the time circle: in principle the background is a supergravity solution, but $T_H$ is of order $\alpha'^{-1/2}$, so at the inverse string scale. 
Nevertheless, the result is known to be correct.

The appearance of a tachyonic mode at a certain ``limiting temperature'' signals the breakdown of the thermal ensemble description, precisely as it happens for the string in ten dimensions.
As far as we know, in eleven dimensions the connection of this breakdown with the exponential growth of the number of states is not proven,\footnote{Note that, concerning the dual field theories, the Hagedorn behavior is not expected to be a property of the planar regime only, as the original Hagedorn paper \cite{Hagedorn:1965st} and lattice studies \cite{Caselle:2015tza} clarify.} so one might prefer to refrain from calling this ``limiting temperature'' the ``Hagedorn temperature''.
Perhaps with an abuse of language, we have chosen to call the ``limiting temperature'' as the ``Hagedorn temperature'', because it reduces to the latter one at small $\eta$ and because it still signals that the used thermodynamic description breaks down.  

Note that, as the author of \cite{Russo:2001vh} points out, we cannot really expect the thermal ensemble to be well defined all the way up to $T_H$, because gravity is not decoupled in that regime. 
But, similarly, the thermal ensemble is not well defined all the way up to $T_H$ also in the more familiar ten-dimensional case, as discussed for example in \cite{Atick:1988si}.
As in that case, we have decided to go on with the investigation, regardless of the existing known problems with the definition of the ensemble.

Finally, as we will mention in section \ref{sec:11d}, 
in the intermediate $\eta$ regime we cannot exclude that our results receive corrections (perhaps connected to the self-gravitating stringy effects mentioned above).
We will anyway analyze this regime, because we expect our results to be at least a part of the full story.

In the second part of the paper, we come back to holographic descriptions in ten dimensions and, using the known formulas from \cite{Urbach:2023npi,Bigazzi:2023hxt}, we calculate the dynamical flavor effects on the Hagedorn temperature in (genuinely) confining theories.\footnote{The behavior in conformal theories on spheres can be generically different from the one described below.}
To be precise, we compute the value of $T_H$ normalized by the (square root of) the string tension.

We consider (almost\footnote{\label{footno}We omit the backgrounds in \cite{Arean:2010hu} from our analysis, since they heavily rely on numerical solutions from which it is difficult to extract clear information for our purposes.}) all the backgrounds we are aware of that are dual to flavored confining theories (with fully backreacting flavors) and that provide a regular description of the IR regime of the dual theories (the one relevant for the calculation of $T_H$).
These backgrounds turn out be a handful.
But, consistently in all the cases, we find out that the effect of flavors, which comes into play at NLO (in $1/\lambda$), is to reduce the value of $T_H$ in the scheme where we take $\lambda$ fixed.
When the flavors are massive, their effect is consistently reduced if the mass is increased.
The reduction effect of $T_H$ due to the flavors is present also for the (normalized) critical temperature for deconfinement $T_c$ when available.
But in this case the reduction is more important, since it is a leading order effect (in $1/\lambda$).
As a result, in the holographic models the flavors tend to separate $T_H$ from $T_c$.

The paper is organized as follows. 
In section \ref{sec:11d} we present our proposal for the effective approach to the calculation of $T_H$ in eleven dimensions.
We consider the analytic form of $T_H$ in the two limiting regimes  $\eta \ll 1$ and $\eta \gg 1$, and explain the structure of the expansion in the whole $\eta$ range. 
We then provide explicit results for the ABJM (section \ref{sec:ABJM}) and WYM models (section \ref{sec:WYM}), where we also discuss the $\theta$-angle dependence.
Then, in section \ref{sec:flavors} we compute the flavor contribution to $T_H$ in confining theories, finding that the Hagedorn temperature is reduced with respect to the unflavored models.
In appendix \ref{sec:appendixCFTs} we collect some results on the flavor effects on $T_H$ in conformal field theories compactified on spheres.

\section{Hagedorn temperature at strong string coupling}
\label{sec:11d} 

In this section we present a proposal for a ``thermal scalar'' eleven-dimensional effective description of the lightest mode of a membrane wrapping both the thermal circle and the eleventh dimension.
The basic idea is that $T_H$ can be calculated in eleven dimensions as the temperature at which a mode on the M-theory membrane wrapped on the thermal circle and on the eleventh dimension becomes tachyonic.
In fact, this has been done in flat space in \cite{Russo:2001vh}.
The result is derived in the large $g_s$ regime, where a quartic interaction term on the world-volume of the membrane can be dropped. 
But, somewhat surprisingly, the result reduces to the known one of the string in the limit $g_s \rightarrow 0$.

Our interest would be to extend this approach to the curved-space case, in order to study e.g.~the $\eta$ corrections in holographic theories such as WYM, ABJM, etc.
One way of doing it would be to write an expansion of the membrane theory around the wrapped configuration, extrapolate its energy to the Hagedorn regime and look at its zero, following the same steps done for the string in \cite{Bigazzi:2022gal,Bigazzi:2023hxt,Bigazzi:2023oqm,Bigazzi:2024biz}.

Another possibility, which we pursue here, is to write down an effective action for the mode which is becoming massless at $T_H$ \`a la Horowitz-Polchinski \cite{Horowitz:1997jc}, but in the eleven-dimensional background instead of the ten-dimensional one.
The reduction of this effective action
to ten dimensions gives the usual thermal scalar effective description.
In this section we start by presenting the proposal.
It provides the value of the Hagedorn temperature $T_H$ for different ranges of the parameter $\eta \sim \lambda^a/N$ ($a>0$).
Then, in sections \ref{sec:expansion}, \ref{sec:runningparameter} we set up the expansion in (the generalization of the parameter) $\alpha'$ which allows to study subleading corrections in $1/\lambda$.
We provide analytic formulae in the two extremal regimes $\eta \ll 1$ and $\eta \gg 1$, and a semi-analytic formula in the remaining regime.

Our results will have the same regime of validity in $g_s \sim \eta$ of the flat space one in  \cite{Russo:2001vh}.
In particular, we ignore the contribution of the quartic interaction term on the membrane.
Nevertheless, in the analogous situation for the string in ten dimensions, quartic world-sheet fluctuations contribute to the results starting at NNLO in the $1/\lambda$ expansion \cite{Bigazzi:2024biz}. 
One could envisage the same to happen for the membrane, in which case our results would be unaffected by the quartic term up to NLO for every value of $\eta$ (i.e.~not just for large $\eta$). These first corrections are thus expected to be totally captured by the zero-mode part of the world-volume fields.


Let us now present our proposal.
The ten-dimensional low-energy effective action for the scalar field $\chi$ in the near-Hagedorn limit in flat space is \cite{Horowitz:1997jc}
\begin{equation}\label{action10d}
 S_{10d} \sim \int d^{10} x \left(\eta^{\mu\nu}\partial_{\mu} \bar\chi \partial_{\nu} \chi + \hat m_{eff}^2 \bar\chi \chi \right) \,,
\end{equation}
with 
\begin{equation}
\hat m_{eff}^2 = \frac{\beta^2 \eta_{\tau\tau} - 8\pi^2\alpha'}{4\pi^2 \alpha'^2} \,,
\end{equation}
where $\tau \simeq \tau + \beta$ is the compact Euclidean time direction, with $\eta_{\tau\tau}=1$. 
The condition $\hat m_{eff}^2=0$ gives the well-known result for the square of the inverse Hagedorn temperature $\beta_H^2 = 8\pi^2\alpha'$.

The natural extension of this action to the flat eleven-dimensional case is
\begin{equation}\label{action11d}
S_{11d} \sim \int d^{11} x \left(\eta^{\mu\nu}\partial_{\mu} \bar\chi \partial_{\nu} \chi +m_{eff}^2 \bar\chi \chi \right) \,,
\end{equation}
with 
\begin{equation}\label{emme}
m_{eff}^2 = \frac{1}{4\pi^2 \alpha'^2} \left[ \beta^2 - \frac{16 \alpha' \zeta(3)}{\sqrt{g_{eff}}} \left(E_{3/2}(g_{eff})-F_{3/2}(g_{eff}) \right) \right]\,.
\end{equation}
The above expression has been constructed in such a way that $m_{eff}^2=0$ gives $\beta_H$ as computed in \cite{Russo:2001vh} by considering a double winding membrane in the eleven-dimensional background. The membrane is supposed to wrap the compact temporal direction and the compact M-theory circle with radius $R_{11}$. The latter is related to the Type IIA coupling $g_s$ as $R_{11} = \sqrt{\alpha'} g_s$.\footnote{In \cite{Russo:2001vh} $g_s$ is called $g_A$.}
Finally, $E_{3/2}, F_{3/2}$ are (pseudo) Eisenstein series
\bea
\label{eisensteinseries}
2\zeta(3)E_{3/2}(g_{eff}) & = & \sum_{(k,m)\neq (0,0)} \frac{\left( \frac{1}{g_{eff}}\right)^{3/2}}{\left|k+\frac{i m}{g_{eff}} \right|^3}\,, \\
2\zeta(3)F_{3/2}(g_{eff}) & = & \sum_{(k,m)\neq (0,0)} (-1)^m\frac{\left( \frac{1}{g_{eff}}\right)^{3/2}}{\left|k+\frac{i m}{g_{eff}} \right|^3}\,, \label{eisensteinseriesF}
\eea 
and
\begin{equation}
g_{eff} =  2\pi\sqrt{\alpha'} g_s /\beta\,.
\end{equation}

\begin{figure}
	\scalebox{0.4}{\includegraphics{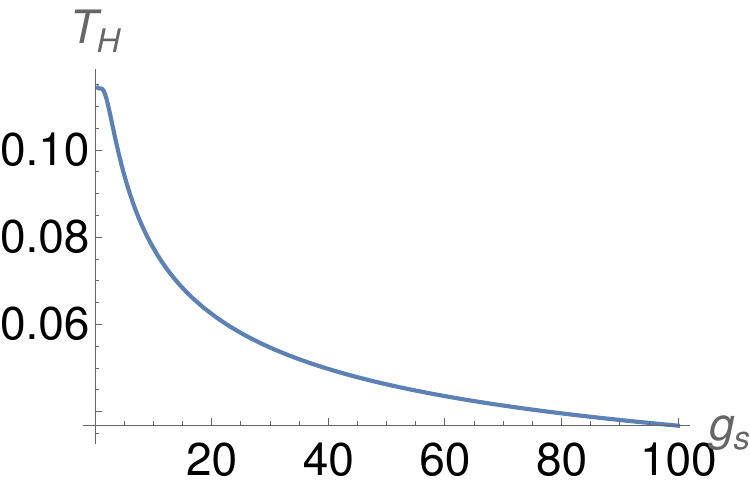}}
	\scalebox{0.4}{\includegraphics{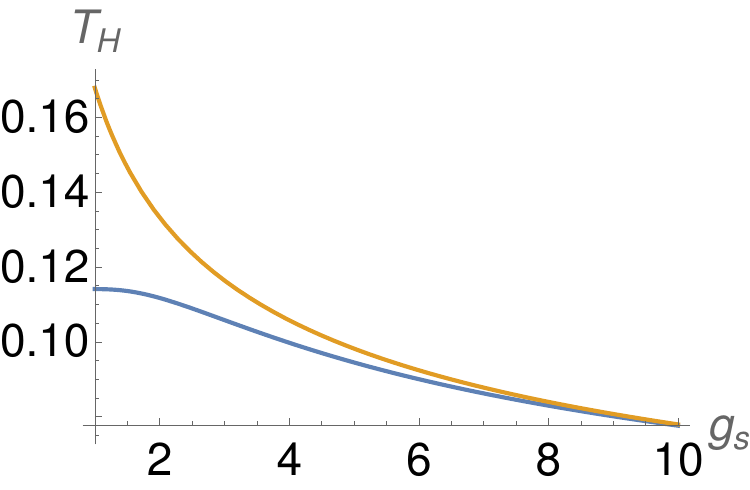}}
	\scalebox{0.4}{\includegraphics{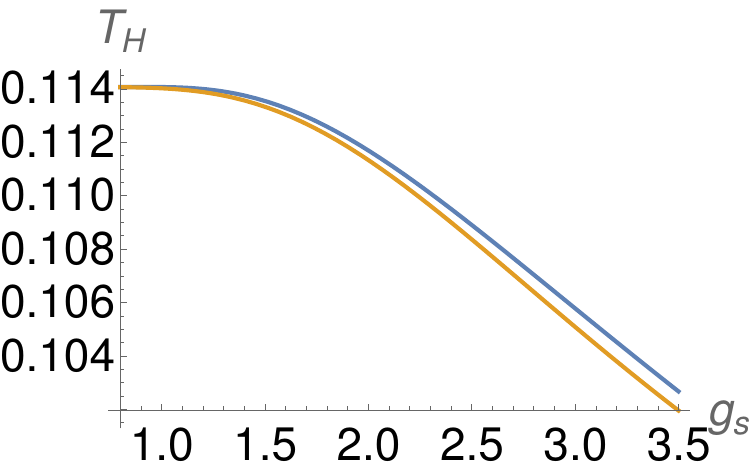}}
	\caption{Left: the flat space Hagedorn temperature in $1/\sqrt{\alpha'}$ units as a function of $g_s$. Center: its comparison with the large $g_s$ analytic result (\ref{largegA}) (orange curve). Right: its comparison with the small $g_s$ analytic result (\ref{smallgA}) (orange curve).}
	\label{fig:flatall}
\end{figure}

The solutions of $m_{eff}^2=0$ for $\beta$ at generic values of $g_s$ can be derived numerically.\footnote{Of course, the numerical accuracy depends on where one cuts the Eisenstein series. In our plots we work with the cut at $|m|=|k|=55$.}
The general form of the Hagedorn temperature as a function of $g_s$ is reported in figure \ref{fig:flatall} on the left (in the plots we set $\alpha'=1$).
For $g_s$ larger than about 10, it can be well approximated by the analytic form (see the center plot of figure \ref{fig:flatall})
\begin{equation}\label{largegA}
	T_H \sim \frac{1}{a \sqrt{\alpha'} \left(2\pi g_s  \right)^{1/3}} \,, \qquad \qquad a = \left[28 \zeta(3) \right]^{\frac13} \sim 3.23 \,, \qquad g_s \gg 1\,.
\end{equation}
For small $g_s$, the formulae in appendices A and B of  \cite{Russo:2001vh} allow to derive its analytic form as
\be \label{smallgA}
T_H \sim \frac{1}{2 \sqrt{2 \alpha'} \pi }\left(1 - \frac{8\cdot 2^{1/4} e^{-\frac{2 \sqrt{2} \pi }{g_s}}}{\pi  \sqrt{g_s}} + ... \right)\,, \qquad g_s \ll 1 \,,
\ee 
and the comparison with the numerical result is reported on the right plot of figure \ref{fig:flatall}. 

Our main point is the generalization of these formulae to the curved space case. 
In order to do so, let us first recall that the tension $T_2$ of the M2-brane is related to the string tension by
\be\label{alphap}
\frac{1}{4\pi^2\alpha'} = T_2 R_{11}\,,
\ee
where $2\pi R_{11}$ is the length of the compact M-theory circle. This relation arises by identifying the wrapped M2-brane action with the string Nambu-Goto one. 

Now, the flat space formula (\ref{emme}) written in terms of M-theory parameters reads
\begin{equation}\label{emmeM}
	m_{eff}^2 = 4\pi^2 T_2^2 R_{11}^2 \beta^2 -8\pi^2\,T_2 R_{11}{\cal F}(g_{eff})\,,
\end{equation}
where
\be\label{geffqui}
g_{eff} = \frac{2\pi R_{11}}{\beta}\,,\qquad {\cal F}(g_{eff}) = \frac{2}{\pi^2} \frac{\zeta(3)}{\sqrt{g_{eff}}} \left(E_{3/2}(g_{eff})-F_{3/2}(g_{eff}) \right)\,.
\ee
Notice that 
\be
{\cal F}(g_{eff}) \rightarrow 1,\,\, (g_{eff}\rightarrow0)\,,\qquad {\cal F}(g_{eff}) \rightarrow \frac{7}{2\pi^2}\zeta(3) g_{eff},\,\, (g_{eff}\rightarrow\infty)\,.
\ee
We consider a generic eleven-dimensional metric, with compact Euclidean time direction $\tau$, in the form
\begin{equation} \label{dictionary}
	ds^2_{11} = G_{\tau\tau} d\tau^2 + ... + G_{AA} (dx_{11} + C_\mu dx^{\mu})^2 = e^{-2\phi/3}ds^2_{10} + e^{4\phi/3} (dx_{11} + C_\mu dx^{\mu})^2 \,,
\end{equation}
so that for the ten-dimensional metric we have $g_{\tau\tau}=G_{\tau\tau} G_{AA}^{1/2}$ (this is true whenever $C_{\tau}=0$, as we are going to assume in the following).
In our conventions the metric functions $G_{AA}$, $G_{\tau\tau}$ and so on do not contain factors of $g_s$.

Extending the expression for $m_{eff}$ to curved space is now immediate: we just perform the replacements
\be
\label{prescri}
R_{11}\rightarrow \sqrt{G_{AA}} \, R_{11}\,,\qquad \beta\rightarrow \beta \, \sqrt{G_{\tau\tau}}\,, \qquad g_{eff}\rightarrow g_{eff}\frac{\sqrt{G_{AA}}}{\sqrt{G_{\tau\tau}}}\,,
\ee
getting
\be\label{newM}
m^2_{eff} = 4 \pi^2 \, T_2^2 R_{11}^2 \sqrt{G_{AA}} \( \sqrt{G_{AA}} \, G_{\tau\tau} \, \beta^2 - \frac{2}{T_2 R_{11}} \, \mathcal F(g_{eff})  \) \, ,
\ee
where, from here on, we keep using the same symbol $g_{eff}$ for the rescaled coupling, that is
\be
\label{geff}
g_{eff} = \frac{2\pi R_{11}}{\beta} \frac{\sqrt{G_{AA}}}{\sqrt{G_{\tau\tau}}} = \frac{\sqrt{\alpha'}}{\beta} \frac{\sqrt{G_{AA}}}{\sqrt{G_{\tau\tau}}} \, 2\pi g_s \, .
\ee
In terms of string theory parameters our formula (\ref{newM}) reads
\be
m_{eff}^2 = \frac{e^{2\phi/3}}{4\pi^2 \alpha'^2}\left[\beta^2 g_{\tau\tau} - 8\pi^2\alpha' {\cal F}(g_{eff})\right]\,,
\ee
so that the reduction of the eleven dimensional action 
\begin{equation} \label{action11dcurved}
S_{11d} \sim \int d^{11} x \sqrt{G} \left(G^{\mu\nu}\partial_{\mu} \bar\chi \partial_{\nu} \chi +m_{eff}^2 \bar\chi \chi \right) \,,
\end{equation}
along the eleventh direction gives
\begin{equation}
	S_{10d} \sim \int d^{10} x \sqrt{g} e^{-2\phi}\left(g^{\mu\nu}\partial_{\mu} \bar\chi \partial_{\nu} \chi +e^{-\frac23\phi} m_{eff}^2 \bar\chi \chi \right) \,,
\end{equation}
which coincides with the generalization of the ten-dimensional action (\ref{action10d}) to curved space (see eq. (3.1) in \cite{Horowitz:1997jc}) once the limit $g_s \rightarrow 0$ is taken in $m_{eff}$.

The rescaling of $g_{eff}$ in equation (\ref{prescri}) can also be understood recalling that the non-perturbative terms in ${\cal F}(g_{eff})$ correspond to D0 instanton contributions in the ($g_{eff}\ll1$) ten-dimensional picture. In particular, one-instanton terms scaling as $e^{-2\pi/g_{eff}}$ are identified with $e^{-S_{D0}}$ terms, where $S_{D0}$ is the on-shell action of a Euclidean D0-brane wrapping the Euclidean time circle
\be
S_{D0}= \frac{1}{g_s l_s} \int d\tau e^{-\phi} \sqrt{g_{\tau\tau}} = \frac{\beta}{R_{11}} e^{-\phi} \sqrt{g_{\tau\tau}}\,.
\ee
Equations (\ref{geffqui}), (\ref{prescri}) consistently give
\be
S_{D0} = \frac{2\pi}{g_{eff}}\,.
\ee

Using the above prescription, the leading order limits for the Hagedorn temperature in the curved case read
\be
\label{leadingbeta}
g_{\tau\tau}(0)\beta_H^2 = 8\pi^2\alpha', \,\, (g_{eff}\rightarrow0)\,,\qquad G_{\tau\tau}(0)^{3/2}\beta_H^3 =\frac{14\zeta(3)}{\pi T_2},\,\, (g_{eff}\rightarrow\infty)\,.
\ee
In the following we are going to study the corrections in $\alpha'$, or equivalently in $1/\lambda$, to these formulae.

\subsection{Perturbative corrections}
\label{sec:expansion}

The equation of motion from the action \eqref{action11dcurved} reads
\be
- \frac{1}{\sqrt{G}} \partial_{\mu} \left[\sqrt{G}  \, G^{\mu\nu} \, \partial_{\nu} \chi \right] +  m^2_{eff} \, \chi = 0 \, .
\ee
We specialize to the case with dependence only on one holographic radial direction $r$ and we assume that the metric is in a form such that the dual to the IR regime of the field theory is located at $r=0$, where there is a Minkowski factor with coordinates $x^i$. 
We consider the ansatz
\be
\chi(x^i , r) = e^{i p^i \eta_{ij}x^j} \, w(r) \, ,
\ee
and take the limit $M^2 = - \eta_{ij} p^i p^j\to0$ (we are assuming that a Wick rotation on the original metric was taken so that one of the directions $x^i$ is timelike).
Moreover, we choose the radial variable $r$ such that the ten-dimensional metric has $g_{rr}=g^{rr}=1$ (so that we can easily compare with the known ten-dimensional results), so that the eleven-dimensional metric has $G^{rr}=e^{+2\phi/3}=G_{AA}^{1/2}$. 
Then the above equation of motion reduces to\footnote{Using $G_{AA}=e^{4\phi/3}$ and $\sqrt{-G}=\sqrt{-g}e^{-8\phi/3}$, one can see that this equation reduces in form to the one in ten dimensions \cite{Bigazzi:2024biz}
	\be\label{eqw10d}
	-\frac12 w'' - \frac12 w' \left(\partial_r \log{\sqrt{-g}}  \right) + w' \phi'  + \frac12 \hat m^2_{eff} w = 0\,,
	\ee
	where we have also to take the $g_{eff} \rightarrow 0$ limit.}
\be\label{eqw}
-\frac12 w'' - \frac12 w' \left(\partial_r \log{\sqrt{-G}}  \right) - \frac14 w' \left(\partial_r \log{G_{AA}}  \right) + \frac{1}{2\sqrt{G_{AA}}} m^2_{eff} w = 0\,.
\ee

At leading order in $\alpha'$ the condition setting the Hagedorn temperature is just $m_{eff}=0$, where the metric functions are evaluated at the ``IR point'' $r=0$ of the geometry where the thermal scalar mode is localized.
Let us call $\sqrt{\alpha'} \beta_L$ the value of the corresponding inverse Hagedorn temperature. This is a function of $g_s$.

At higher orders in $\alpha'$ we assume that the metric admits an expansion in even powers of $r$ and has a $(d-1)$-dimensional shrinking cycle (as usual in backgrounds dual to confining theories), such that
\begin{subequations}
\label{Gexpansion}
\begin{align}
& \sqrt{-G} = const \cdot r^{d-1}(1+G_2 \, r^2+ \ldots) \,, \\
& G_{\tau\tau} = G_{0}^{(0)}(1+G_{2}^{(0)} r^2 + G_{4}^{(0)} r^4 + \ldots)\,, \\
& G_{AA} = G_{0}^{(A)}(1+G_{2}^{(A)} r^2 + G_{4}^{(A)} r^4 +\ldots)\,, \\
&G^{rr} = G_0^{(r)} \(1- G^{(r)}_2 \, r^2 + \ldots\) \, .
\end{align}
\end{subequations}
We also write the expansion for the inverse Hagedorn temperature as
\be 
\beta^2_H = \alpha' \beta_L^2 + \alpha'^{3/2} \beta_{NL}^2 + \alpha'^2 \beta^2_{NNL} + ...
\ee 
It is worth emphasizing that the actual expansion parameter is dependent on the regime of $g_s$ and coincides with $\alpha'$ only at $g_s \ll 1$.
The actual results for the coefficients $\beta_{L,NL,...}$ will carry this information, as we will see.
This issue will be discussed in details in section \ref{sec:runningparameter}, where we will write an expansion parameter valid for every value of $g_s$.
Finally, let us stress that we use the parameter $\alpha'$ for notational convenience.
All the eleven-dimensional results can be written in terms of the eleven-dimensional quantities $T_2, R_{11}$ by means of relation (\ref{alphap}).

In the next sections we will explicitate the expressions of $\beta_H^2$ in different regimes of $g_s$.
We will mainly focus on the LO and NLO, since at NNLO the results are incomplete, missing the contribution of the massive membrane modes.

\subsubsection{Weak coupling limit}
We have already seen that the proposal gives the known results in the strict $g_{eff}\rightarrow 0$ limit, since the eleven-dimensional action reduces to the ten-dimensional one.
The formalism allows to calculate analytically the first non-perturbative corrections at small $g_{eff}$.
In fact, the first such corrections are exponentially suppressed with $g_{eff}$.
As we have seen, in ten dimensions they can be understood as  instantonic contributions from Euclidean D0-branes wrapped on the thermal cycle.

By expanding the (pseudo) Eisenstein series, one finds in this limit
\be \label{meffweak} 
\frac12 m^2_{eff} = \frac{G_{AA}^{1/2}}{8\pi^2\alpha'^2} \left[ G_{\tau\tau}G_{AA}^{1/2}\, \beta^2 - \alpha' \left(8 \pi ^2 +  \frac{128 \pi  e^{-\frac{2 \pi }{ g_{eff}}}}{\sqrt{ g_{eff}}}\right) + {\cal O}\left(e^{-\frac{4 \pi }{ g_{eff}}}\right) \right]\,.
\ee 
Thus, at leading order in $\alpha'$ and small $g_{eff}$ the inverse Hagedorn temperature can be determined from the condition $m_{eff}=0$ as
\be \label{betaLweak}
\beta_L^2 = \frac{8\pi^2}{G_{0}^{(0)}(G_{0}^{(A)})^{1/2}} \left[1 + \frac{2^{7/2}  e^{-\frac{1}{\eta}}}{\pi^{3/2}\sqrt{\eta}} \right]\,, 
\ee 
where we have defined the new expansion parameter
\be \label{eta}
 \eta \equiv  \frac{(G_0^{(A)})^{3/4}}{2\sqrt{2}\pi} g_s \,.
\ee

In order to analyze the NLO corrections\footnote{In the notation we are adopting the terms ``LO'', ``NLO'', etc. always refer to the (generalized) $\alpha'$ expansion, while for the expansions in $\eta$ we will be explicit.} in $\alpha'$ we expand equation (\ref{eqw}) with $m_{eff}$ given by (\ref{newM}). 
We get the equation for a $d$-dimensional harmonic oscillator
\be\label{harmonicweak}
- \frac12 \, w'' - \frac12 \, \frac{d-1}{r} \, w' + \frac12 \omega^2 \, r^2 w = E_0  w\,,
\ee
with frequency
\be \label{freq}
\frac12 \omega^2 = \frac{1}{8\pi^2 \alpha'} \left[G_0^{(0)} (G_0^{(A)})^{1/2} \left(G_2^{(0)}+\frac12 G_2^{(A)}\right) \beta_L^2 + (G_2^{(A)}- G_2^{(0)}) \frac{2^4 \sqrt{2\pi}  e^{-\frac{1}{\eta}}}{\sqrt{\eta}}\left(1-\frac{2}{\eta}\right) \right]\,,
\ee 
and ground-state energy
\be  \label{gro}
E_0 = \frac{d}{2}\omega = - \frac{G_0^{(0)} (G_0^{(A)})^{1/2} }{8\pi^2 (\alpha')^{1/2}} \beta^2_{NL} \left[1 - 8 \, \frac{e^{-\frac{1}{\eta}}}{(2\pi)^{\frac32} \sqrt{\eta}}\left(1-\frac{2}{\eta}\right) \right]\,.
\ee 
In order to obtain these formulae we have first expanded in $\alpha'$ the exact function ${\cal F}(g_{eff})$ evaluated at LO, and then we have taken the leading contributions at small $\eta$. 
That is, we have not performed a double expansion in $\eta$ (first) and $\alpha'$ (after), in order to avoid problems with the non-analytic terms at small $\eta$.\footnote{In particular, we are not perturbing around the one-instanton configuration.} 
This will be hopefully clear in the general discussion in section \ref{sec:runningparameter}, where the perturbative expansion (with expansion parameter called $s$ there) is defined around any given value of $\eta$.  

From (\ref{freq}), (\ref{gro}) it follows that $\beta_{NL}$ is
\begin{eqnarray}
\label{NLweak}
&& \beta_{NL}^2 = -\frac{4\sqrt{2}\pi^2 d}{G_0^{(0)} (G_0^{(A)})^{1/2}} \sqrt{G_2^{(0)}+\frac12 G_2^{(A)}}\Biggl[ 1 +   \frac{3\cdot 2^3   e^{-\frac{1}{\eta}}}{(2\pi)^{3/2}\sqrt{\eta}}
\frac{(5G_2^{(0)}+4G_2^{(A)})}{(6G_2^{(0)}+3G_2^{(A)})}  \cdot \nonumber \\
&& \qquad  \qquad \qquad \qquad \qquad \qquad \qquad  \qquad \qquad  \qquad \cdot \left(1-\frac{(2G_2^{(0)}+4G_2^{(A)})}{\eta(5G_2^{(0)}+4G_2^{(A)})} \right)
\Biggr]\,.
\end{eqnarray}

\subsubsection{Strong coupling limit}

Let us now look at the opposite limit $g_{eff}\rightarrow \infty$.
We have
\be  
\frac12 m^2_{eff} = \frac{G_{AA}^{1/2}}{8\pi^2\alpha'^2} \left[ G_{\tau\tau}G_{AA}^{1/2}\, \beta^2 - \alpha' \left( a^3 g_{eff} +\frac{16 \pi^2}{3} \frac{1}{g_{eff} }\right) + {\cal O}\left(e^{-2 \pi g_{eff}}\right)\right]\,.
\ee 
Remember that $g_{eff}$ depends on $\beta$.
Thus, up to next-to-leading order in $1/g_{eff}$, $m_{eff}=0$ gives
\be \label{betaLstrong}
\beta_L^2 = \frac{a^2 2^{5/3}\pi^{4/3}}{G^{(0)}_0 (G_0^{(A)})^{1/2}} \eta^{2/3} \left[ 1 + \frac{2^{5/3}}{9 a \pi^{2/3}}\frac{1}{\eta^{4/3}} \right] \,.
\ee 

At NLO in $\alpha'$, the expansion of the metric functions and $\beta$ gives for the equation of motion (\ref{eqw})
\be
 - \frac12 \, w'' - \frac12 \, \frac{d-1}{r} \, w' + \frac12 \omega^2 \, r^2 w = E_0  w\,.
\ee
This is again a $d$-dimensional harmonic oscillator with frequency
\bea 
\frac12 \omega^2 & = & \frac{G_0^{(0)} (G_0^{(A)})^{1/2} }{8\pi^2 \alpha'} \Biggl[ \left(G_2^{(0)}+\frac12 G_2^{(A)}\right) \beta_L^2 - \frac{a^3 2^{3/2}\pi^2 \eta }{ (G^{(0)}_0)^{3/2}(G^{(0)}_A)^{3/4}}(G_2^{(A)}- G_2^{(0)}) \frac{1}{\beta_L} + \nonumber \\
&& \qquad \qquad\qquad- \frac{2^{3/2}}{3} \frac{G_2^{(0)}- G_2^{(A)}}{2(G^{(0)}_0)^{1/2} (G^{(A)}_0)^{1/4}}\frac{\beta_L}{\eta} \Biggr]\,,
\eea 
and ground-state energy
\be  
E_0 = \frac{d}{2}\omega = - \frac{G_0^{(0)} (G_0^{(A)})^{1/2} }{8\pi^2 \alpha'^{1/2}} \beta^2_{NL}\left[ 1 + \frac{a^3 2^{3/2}\pi^2 \eta }{ (G^{(0)}_0)^{3/2}(G^{(0)}_A)^{3/4}}\frac{1}{\beta_L^3} - \frac{2^{1/2}}{3} \frac{1}{(G^{(0)}_0)^{1/2} (G^{(A)}_0)^{1/4}}\frac{1}{\beta_L\eta}   \right]\,,
\ee 
which give
\be
\label{NLstrong}
\beta_{NL}^2 = - \frac{d\, 2^{7/3} \pi^{5/3} a (G_2^{(0)})^{1/2} }{\sqrt{3}G_{0}^{(0)}(G_0^{(A)})^{1/2} } \eta^{1/3} \left[ 1 + \frac{ 2^{2/3}(2 G_2^{(0)}+ G_2^{(A)})}{9 \pi^{2/3} a G_2^{(0)}} \frac{1}{\eta^{4/3}} \right]\,.
\ee
Following the conventions in \cite{Russo:2001vh}, that is
\be \label{lplanck}
l_P = \sqrt{\alpha'}(2\pi g_s)^{1/3} = \sqrt{\alpha'} \frac{2^{5/6} \pi^{2/3}}{(G_0^{(A)})^{1/4}} \eta^{1/3} \,,
\ee 
one gets that $\beta_H^2$ is proportional to $l_P^2$, with corrections which come in a double expansion in $1/\eta^{4/3}$ and $\alpha'/l_P$.

The expansion in $\alpha'/l_P$ is confirmed by the result at leading order in $g_s$ and NNLO in $\alpha'$.
Even though we can expect further corrections at this order, we are going to present the results coming from the effective approach.
As usual, this calculation involves considering the first order perturbation to the unperturbed Hamiltonian $H_0$, 
\be \label{H1strong}
H_1 w = E_1 w \,, 
\ee  
where $H_1$ and $E_1$ are extracted from the expansion of the equation of motion at the next order.
The expressions are a bit messy, but the structure is clear,
\be
H_1  =  \alpha\, r^2 (H_0 - E_0) + \Delta H_1 \,,
\ee
where $\alpha$ is a known expressions we will not need and
\be\label{deltah1}
\Delta H_1  =    -A\, r\, \partial_r + B r^2 + C r^4 \,,
\ee
with
\bea 
A & = &  \left[G_2 +\frac12 G_2^{(A)}\right] \,,\\
B & = &   \frac{G_0^{(0)}(G_0^{(A)})^{1/2}}{8\pi^2 \sqrt{\alpha'}} \beta^2_{NL} \left[G_2^{(0)} +\frac12 G_2^{(A)} +\frac{a^3 2^{1/2} \pi^2 \eta }{ \beta_L^3 (G_0^{(0)})^{3/2}(G_0^{(A)})^{3/4}}(G_2^{(A)}-G_2^{(0)}) \right] \,, \\
C & = & \frac{G_0^{(0)}(G_0^{(A)})^{1/2}}{64\pi^2 \alpha'} \beta^2_{L}  \Biggl[  4G_2^{(0)}G_2^{(A)} -(G_2^{(A)})^2 +8 G_4^{(0)} + 4G_4^{(A)}  + \\
&& \qquad \qquad - \frac{a^3 2^{5/2} \pi^2 \eta }{ \beta_L^3 (G_0^{(0)})^{3/2}(G_0^{(A)})^{3/4}} \left( 3(G_2^{(0)})^2 -2G_2^{(0)}G_2^{(A)} -(G_2^{(A)})^2 -4 G_4^{(0)} - 4G_4^{(A)}  \right) \Biggr] \,. \nonumber
\eea 
Moreover
\be \label{E11}
E_1 =- \frac{G_0^{(0)}(G_0^{(A)})^{1/2}}{8\pi^2} \left[ \beta^2_{NNL}   - \frac{a^3 \pi^2 \eta }{ 2^{1/2} \beta_L (G_0^{(0)})^{3/2}(G_0^{(A)})^{3/4}} \left(3\frac{\beta^4_{NL}}{\beta^4_{L}} - 4\frac{\beta^2_{NNL}}{\beta^2_{L}} \right) \right]\,.
\ee

In perturbation theory
\be \label{eqforE1}
E_1 = \langle w_0 | H_1 | w_0\rangle = \langle w_0 | \Delta H_1 | w_0\rangle \,,
\ee 
where $w_0$ is the (normalized) ground state of the Harmonic oscillator Hamiltonian $H_0$ with energy $E_0$, i.e.
\be 
w_0= \alpha_0 e^{-\frac{\omega r^2}{2}}\,, \qquad 1=\int_0^{\infty} dr r^{d-1} |w_0|^2\,.
\ee 
The calculation of (\ref{eqforE1}) with $\Delta H_1$ in (\ref{deltah1}) gives the result
\be \label{E12}
E_1 = \frac{d}{2}A + \frac{d}{2}\frac{B}{\omega} + \frac{d(d+2)}{4}\frac{C}{\omega^2}\,.
\ee 
Equating (\ref{E11}) with (\ref{E12}) allows to extract
\bea \label{NNLstrong}
\beta^2_{NNL} & = & \frac{\pi^2 d }{18 G_0^{(0)} (G_0^{(A)})^{1/2} G_2^{(0)}}  \Biggl[ -48 G_2 G_2^{(0)}+ 3(2+9d) (G_2^{(0)})^2 +6(d-6) G_2^{(0)} G_2^{(A)} \nonumber \\
&& \qquad \qquad - 4 (d+2) \left(3 G_4^{(0)}+2 G_4^{(A)}\right) \Biggr]\,,
\eea
which is independent on $\eta$, that is on $g_s$.

\subsection{Running expansion parameter}
\label{sec:runningparameter}

In the previous sections, we studied the problem at weak and strong coupling separately. The question that naturally arises is: what happens in the intermediate regime? First of all, we need to understand whether or not it makes sense to address this problem.

As we already mentioned at the beginning of this section, this work relies on the flat space computation by Russo in \cite{Russo:2001vh}. There, he neglected an interaction term in the membrane Hamiltonian. This is perfectly reasonable at strong coupling, since it scales as $g_s^{-2}$ (see his formula (4.3)). However, it may not be the case far from this regime.

The situation is quite interesting from the point of view of the Hagedorn temperature in curved space. Indeed, the interaction Hamiltonian neglected by Russo is quartic in the world-volume fluctuations. Then, by analogy with the string theory computation, it should affect the Hagedorn temperature starting from the NNLO in the expansion parameter (see \cite{Bigazzi:2024biz}). Therefore, the NLO correction to $\beta_H$ should be robust at any value of the coupling $g_s$.

All in all, this reasoning should justify the effort made in trying to answer the above question. Here, we propose a NLO computation which should be valid for all possible values of $g_s$. 

To begin with, let us look again at the expansion around $r=0$ of the eleven dimensional metric in \eqref{Gexpansion}. At the Hagedorn point, the masses of the bosonic membrane fluctuations are thus expected to be
\be
\mu = \frac{\beta_H}{2\pi} \frac{\sqrt{G_0^{(0)}}}{l_{11}} \, ,
\ee
where $l_{11}\sim\(G^{(0)}_2\)^{-1/2}$ is the length scale of the eleven-dimensional background. 

Moreover, as usual, the leading order inverse Hagedorn temperature $\beta_{H, LO}$ is fixed by the condition $m^2_{eff} = 0$. In other words, it is the solution of
\be
\label{betaHLO}
\beta^2 = \frac{1}{G_0^{(0)}(G_0^{(A)})^{1/2}} \frac{2}{T_2 R_{11}} \, \mathcal F\(\frac{2\pi R_{11}}{\beta} \sqrt{\frac{G_0^{(A)}}{G_0^{(0)}}}\)\, .
\ee
This statement holds for any possible value of $g_s$. In principle, $\mathcal F$ is defined as in \eqref{geffqui} with $g_{eff}$ modified as in \eqref{geff}. If we focus on a particular regime of $g_s$, we can give an approximated version of $\mathcal F$. For instance, keeping the first subleading corrections in $g_{eff}$ (and so in $g_s$, see \eqref{geff}), in the asymptotic regimes it reads
\be \label{calF}
{\cal F}(g_{eff}) \approx
\begin{cases}
\displaystyle 1 + \frac{16}{\pi} \frac{1}{\sqrt{g_{eff}}} e^{-\frac{2\pi}{g_{eff}}}\, , & g_{eff} \ll 1 \, , \\
\displaystyle \frac{7}{2\pi^2} \, \zeta(3) \, g_{eff} + \frac23 \frac{1}{g_{eff}} \, , & g_{eff} \gg 1 \, .
\end{cases}
\ee

We are thus tempted to introduce the parameters\footnote{Notice that $h$ takes the leading order value of $g_{eff}$.}
\be
\label{runningepsilon}
\varepsilon = \frac{2}{T_2 R_{11}} \, \mathcal F(h) \, , \qquad h = \frac{2\pi R_{11}}{\beta_{H,LO}} \sqrt{\frac{G_0^{(A)}}{G_0^{(0)}}} \, , 
\ee
so that
\be
\mu \sim \sqrt{\varepsilon} \, .
\ee
Nevertheless, what appears in $m^2_{eff}$ is the combination $4\pi^2 T_2^2 R_{11}^2 \varepsilon$.
Furthermore, in the notations of \cite{Russo:2001vh}, the radial zero-mode scales as
\be
r \sim \frac{\sqrt{\alpha'}}{\sqrt{\mu}} \sim (4\pi^2 T_2^2 R_{11}^2 \varepsilon)^{-1/4} \, .
\label{scalingr}
\ee

In other words, the equations suggest the existence of the length scale
\be
\label{s}
s = (4\pi^2 T_2^2 R_{11}^2 \varepsilon)^{-1/2} \, .
\ee
Let us stress that this is \emph{not} an implicit equation for $s$. Indeed, neither was the definition of $\varepsilon$ in \eqref{runningepsilon}. Once one computes the value of $\beta_{H,LO}$ by means of \eqref{betaHLO}, the value of $s$ is automatically fixed thanks to the above definition. 

Of course, $s$ depends on the value of $g_s$ at which we decide to make computations. For instance, taking into account the first subleading corrections in $g_s$ (see \eqref{calF}, on the right), it reads\footnote{At each (different) value of $g_s$, the parameter $s$ can be expressed as a (different) combination of the field theory parameters $N$ and $\lambda$.}
\be \label{asyms}
s \approx 
\begin{cases}
\displaystyle \sqrt{\frac{\alpha'}{2}} \( 1 - \frac{16}{(2\pi)^{3/2}} \frac{e^{-\frac{1}{\eta}}}{\sqrt{\eta}} \) \, , & \eta \ll 1 \, , \\[2ex]
\displaystyle b \, \frac{\ell_p^2}{R_{11}} \( 1 - \frac{1}{9 a } \, \frac{2^{2/3}}{\pi^{2/3}} \, \frac{1}{\eta^{4/3}} \)\, , & \eta \gg 1 \, ,
\end{cases}
\ee
where
\be \label{parameters}
b^{-1}= a \, (G_0^{(A)})^{1/4} \, , 
\ee
and $a$ and $\eta$ have been defined respectively in \eqref{largegA} and \eqref{eta}.
Notice that the appearance of $R_{11}$ is reasonable. Indeed, the (double) winding membrane is wrapped both on the thermal circle and on the compact eleventh-direction.

It is thus natural to parameterize the Hagedorn temperature as
\be
\label{betaansatz}
4 \pi^2 \, T_2^2 R_{11}^2 \, \beta_H^2 = \frac{s^{-2}}{G_0^{(0)} (G_0^{(A)})^{1/2}} \( 1 + c_0 \, s + c_1 \, s^2 + \ldots \) \, .
\ee
Indeed, since, as already pointed out in (\ref{scalingr}), the radial zero-mode scales as
\be
\label{zeromodesscaling}
r \sim \sqrt{s} \, ,
\ee
with respect to the length scale $s$ defined in (\ref{s}), it is easy to expand the equation of motion \eqref{eqw}
in powers of $s$, identifying each order of the perturbative series precisely. 
For future convenience, we will treat the expansions in power of $r$ and $s$ as independent. Nevertheless, thanks to the scaling in \eqref{zeromodesscaling}, they are equivalent. 

Keeping this in mind, let us focus on the expansion of $m^2_{eff}$. In particular, notice that, by definition,
\be
s^2 \, 8 \pi^2 \, T_2 R_{11} \, \mathcal F(g_{eff}) = s^2 \, 4 \pi^2 \, T_2^2 R_{11}^2 \frac{2}{T_2 R_{11}} \, \mathcal F(g_{eff}) \sim \mathcal O (s^0) \, .
\ee
Then, the above quantity can be expanded by means of a double Taylor series around $(r,s)=(0,0)$, that is
\be
s^2 \, 8 \pi^2 \, T_2 R_{11} \, \mathcal F(g_{eff}) = \sum_{n,m=0}^{+\infty} F_{(2n,m)} \, r^{2n} \, s^m \, ,
\ee
where we have defined
\be\label{F2}
F_{(2n,m)} = \left .\frac{1}{(2n)! \, m!} \, \frac{\partial^{2n+m}}{\partial r^{2n} \partial s^m} \(s^2 \, 8 \pi^2 \, T_2 R_{11} \, \mathcal F(g_{eff})\) \right |_{r,s=0} \, .
\ee
Each term in the series is $\mathcal O (r^{2n} s^m) = \mathcal O(s^{n+m})$.

Following the conventions of section \ref{sec:expansion} (see \eqref{Gexpansion}), the final expression for the effective square mass is thus
\be
\label{finalmeff2}
m^2_{eff} = \frac{(G_0^{(A)})^{1/2}}{s^2} \[ 1 + \( \frac12 G_2^{(A)} + G_2^{(0)}\) r^2 + c_0 \, s + \ldots - \sum_{n,m=0}^{+\infty} F_{(2n,m)} \, r^{2n} \, s^m \]\,.
\ee
Remember that here we aim to make an order by order computation. In particular, our goal is the NLO coefficient $c_0$ in the expansion \eqref{betaansatz}. 
So, let us focus on the leading contribution in the $s$-expansion of the equation of motion, that is\footnote{Notice that, by definition, $F_{(0,0)}=1$. So its contribution cancels out with the $1$ in the square bracket of formula \eqref{finalmeff2}.}
\be
- \frac12 w'' - \frac12 \frac{d-1}{r} w' + \frac{1}{2s} \(c_0 - F_{(0,1)}\) w + \frac{1}{2s^2} \( \frac12 G_2^{(A)} + G_2^{(0)} - F_{(2,0)}\) r^2 \, w =0 \, .
\ee

Introducing the zeroth-order variable
\be
x=\[\frac{1}{s^2} \(\frac12 G_2^{(A)} + G_2^{(0)} - F_{(2,0)}\)\]^{1/4} r \, ,
\ee
the equation of motion takes the form of the eigenvalue problem for a $d$-dimensional harmonic oscillator with unit frequency, that is
\be
-\frac12 \frac{\partial^2}{\partial x^2} w - \frac12 \, \frac{d-1}{x} \, \frac{\partial}{\partial x} w + \frac12 \, x^2 \, w = E_0 \, w \, ,
\ee
where
\be
E_0 = - \frac{1}{2} \(c_0 - F_{(0,1)}\) \(\frac12 G_2^{(A)} + G_2^{(0)} - F_{(2,0)}\)^{\hspace{-4pt}-1/2} \, .
\ee

If we want the problem to admit a normalizable solution,
that is
\be
w \sim e^{-\frac{x^2}{2}} \, ,
\ee
we have to require
\be
E_0 = \frac{d}{2} \, .
\ee

All in all, we conclude that
\be\label{c0}
c_0 = F_{(0,1)} - d \, \(\frac12 G_2^{(A)} + G_2^{(0)} - F_{(2,0)}\)^{\hspace{-4pt}1/2} \, .
\ee
By construction, $F_{(0,1)}$ should be linear in $c_0$. Therefore, the one above is an implicit equation for $c_0$.
The problem is thus reduced to the computation of the coefficients $F_{(0,1)}$ and $F_{(2,0)}$, by means of their definition in \eqref{F2}.

To begin with, we prepare the ground with a whole series of observations. First of all, looking at the perturbative expansion in \eqref{betaansatz}, it follows that 
\be
\label{twofold}
\beta_{H,LO}^2 (1+c_0\,s) = \beta_H^2 = \frac{s^{-2}}{4\pi^2 T_2^2 R_{11}^2} \frac{(1+c_0\,s)}{G_0^{(0)}(G_0^{(A)})^{1/2}} \, .
\ee
Then, despite $\beta_{H,LO}$ is an $s$-independent c-number fixed by \eqref{betaHLO}, if necessary we can write it in terms of $s$. This can help us if we have to absorb positive powers of $s$, as in the definition of $F_{(0,1)}$ and $F_{(2,0)}$ in \eqref{F2}.

Furthermore, from the expressions of the (pseudo) Eisenstein series in \eqref{eisensteinseries}, \eqref{eisensteinseriesF}, $\mathcal F(g_{eff})$ takes the form
\be
\label{fullglorycalF}
\mathcal F(g_{eff}) = \frac{2}{\pi^2 g^2_{eff}} \sum_{\substack{k,n\in\mathbb Z\\k \neq 0}} \( k^2 + \frac{(2n-1)^2}{g_{eff}^2} \)^{\hspace{-4pt}-3/2} \, .
\ee

All in all, we can write\footnote{\label{strategy}To get there, we have used the first equality in \eqref{twofold} within the series, while we have used the second one in the overall factor. Moreover, combining the implicit equation for $\beta_{H,LO}$ in \eqref{betaHLO} with the definition of $h$ in \eqref{runningepsilon}, it follows that $h^2=4\pi^2\eta^2/\mathcal F(h)$. The latter has been used to massage the overall factor.}
\be \label{staylor}
\left . s^2 8\pi^2 \, T_2 R_{11} \, \mathcal F (g_{eff}) \right |_{r=0} = \frac{2 (1+c_0\,s)}{\mathcal F(h) \pi^2 h^2} \sum_{\substack{k,n\in\mathbb Z\\k \neq 0}} \( k^2 + \frac{(2n-1)^2}{h^2} (1+c_0\,s) \)^{\hspace{-4pt}-3/2} \, .
\ee
By definition we have
\be \label{almostfinalF01}
F_{(0,1)} = \left . \partial_s \(s^2 8\pi^2 \, T_2 R_{11} \, \mathcal F (g_{eff})\) \right |_{r,s=0} \, .
\ee
Then, plugging \eqref{staylor} into the above \eqref{almostfinalF01}, it is easy to show that
\be \label{F01}
F_{(0,1)} = - \frac12 \, h \, \frac{\partial}{\partial h} \log \mathcal F (h) \,  c_0 \, .
\ee

Now, let us focus on $F_{(2,0)}$, which is defined as
\be
F_{(2,0)} = \frac12 \left . \partial^2_r \(s^2 8\pi^2 \, T_2 R_{11} \, \mathcal F (g_{eff})\) \right |_{r,s=0} \, .
\ee
Combining the definitions of $\epsilon$ and $s$ respectively in \eqref{runningepsilon} and \eqref{s} and using the chain rule, it follows that\footnote{Notice that $g_{eff}$ admits, by assumption, an expansion in even power of $r$ around $r=0$. Therefore, its first derivative vanishes at $r=0$.}
\be
F_{(2,0)} = \frac12 \left . \partial^2_r g_{eff} \right |_{r,s=0} \frac{\partial}{\partial h} \log\mathcal  F(h) \, .
\ee
Looking at the definition of $g_{eff}$ in \eqref{geff}, we finally get
\be \label{F20}
F_{(2,0)} = \frac12 \(G_2^{(A)} - G_2^{(0)}\) \, h \, \frac{\partial}{\partial h} \mathcal \log \mathcal F (h) = -\frac{F_{(0,1)}}{c_0} \(G_2^{(A)} - G_2^{(0)}\)  \, .
\ee

We are now ready to write a closed formula for $c_0$ as
\be
\label{runningc0}
\boxed{
c_0 = - d \, \(1 + \frac12 \, h \, \frac{\partial}{\partial h} \log{\mathcal F (h)}\)^{\hspace{-4pt}-1} \[\frac12 G_2^{(A)} + G_2^{(0)} - \frac12 \(G_2^{(A)} - G_2^{(0)}\) \, h \, \frac{\partial}{\partial h} \mathcal \log \mathcal F (h)\]^{\hspace{-2pt}1/2}
} \, .
\ee

Let us stress that we never specified a regime for $g_s$ so far. We just expanded $g_{eff}$ around its leading order value $h$ in the $(r,s)$-expansion, regardless of the magnitude of $h$ (e.g.~see \eqref{staylor}). What appears here is the expression for $\mathcal F$ in \eqref{fullglorycalF}, which formally depends on $h \sim \mathcal O(s^0)$. The expression for $c_0 \sim \mathcal O (s^0)$ in \eqref{runningc0}, and in turn also the formal expansion of $\beta_H$ in \eqref{betaansatz}, is thus valid for any value of the coupling. As a consequence, this formula allows to catch at the same time the behavior of the Hagedorn temperature for every value of $g_s$ and $s$ such that $c_0 \, s < 1$. Otherwise, the perturbative expansion would break down. 

Anyway, due to the form of the function $\mathcal F$, away from the limiting values of $g_s$, $c_0$ must be evaluated numerically. Notice that, once we fix the background metric, $c_0$ depends just on $h$. Moreover, using formulas (\ref{betaHLO}) and (\ref{runningepsilon}) (or combining the definition of $\eta$ in \eqref{eta} with the results provided in footnote \ref{strategy}), we can assign a value to $g_s$ given any $h$ as
\be \label{gsh}
g_s=\(G_{0}^{(A)}\)^{\hspace{-2pt}-3/4} \, h \, \sqrt{2 \,\mathcal F (h)} \, .
\ee
Therefore, it is very convenient to plot
\be \label{plottable}
\frac{T_H}{T_{H,LO}} = \(1+ c_0(g_s) \, s\)^{-\frac12} \approx 1 - \frac12 c_0(g_s) \, s \, ,
\ee
where $g_s$ among brackets stands for the solution of \eqref{gsh} as $h$ varies. Notice that $s$ is meant to be small at any value of $g_s$ (e.g., see \eqref{asyms}). 

If we specify the regime of $g_s$, then we can provide something more explicit. Of course, the expression of the (running) expansion parameter $s$ in terms of the string or Planck scale and $g_s$ relies case by case on the regime we want to investigate, as we already discussed before (see \eqref{asyms}). Therefore, the recipe to be followed is:
\begin{enumerate}
\item identify the regime of $g_s$ of interest;
\item select the approximated version of $\mathcal F$ consistently with that regime;
\item compute the value of $\beta_{H,LO}$ as the solution for \eqref{betaHLO} (this could be an implicit equation) and $c_0$ according to \eqref{runningc0};
\item provide the NLO value of $\beta_H$ as $\displaystyle \beta^2_H = \beta^2_{H,LO} \(1 + c_0 \, s\)$.
\end{enumerate}

Let us apply this procedure to a couple of relevant examples. Relying on the Fourier expansions in appendices A and B of \cite{Russo:2001vh}, it turns out that $\mathcal F$ can be rephrased as
\begin{align}
\mathcal F (h) &= 1 + \frac{32}{\pi \, h} \sum_{n,m=1}^{+\infty} \frac{n}{2m-1} K_1 \(\frac{2\pi n (2m-1)}{h}\) \nonumber \\
&= \frac{7}{2\pi^2} \, \zeta(3) \, h + \frac23 \, \frac1h + \frac{16}{\pi}  \sum_{n \in \frac12 \mathbb N^+} \sum_{m=1}^{+\infty} (-1)^{2n} \frac{n}{m} K_1\(2\pi \, n \, m \, h\) \, .
\end{align}
Here, $\frac12 \mathbb N^+=\{\frac12,1,\frac32,2,...\}$ denotes the set given by the union of the positive integers and half-integers. Let us stress that these expressions are exact in $h$, they are not perturbative expansions around a specific value of $h$. In other words, they are valid for any value of $h$, even if the first (second) one lends itself to an expansion for small (large) $h$. The same remarks hold for the derivatives of $\mathcal F$. Indeed, from well-known properties of the Bessel functions, we can express them in terms of non-perturbative series as
{\small{
\begin{eqnarray}
\frac{\partial \mathcal F (h)}{\partial h} & = & -\frac{32}{\pi \, h^2} \sum_{n,m=1}^{+\infty} \Biggl[ \frac{n}{2m-1} K_1 \(\frac{2\pi n (2m-1)}{h}\)+   \\
&& \qquad  \qquad \qquad \qquad  - \frac{\pi\,n^2}{h}\[K_0 \(\frac{2\pi n (2m-1)}{h}\)+K_2 \(\frac{2\pi n (2m-1)}{h}\)\] \Biggr]  \nonumber \\
& = & \frac{7}{2\pi^2} \, \zeta(3) - \frac23 \, \frac{1}{h^2} -16 \sum_{n \in \frac12 \mathbb N^+} \sum_{m=1}^{+\infty} (-1)^{2n} \, n^2 \[K_0\(2\pi \, n \, m \, h\)+K_2\(2\pi \, n \, m \, h\)\] \, . \nonumber
\end{eqnarray}
}}

All in all, if we are interested just in the first subleading corrections in $g_s$, at weak (but not so weak) coupling, everything goes as if
\be
\mathcal F (h) \approx 1 + \frac{16}{\pi} \frac{e^{-\frac{2\pi}{h}}}{\sqrt{h}} \, , \quad h \ll 1 \, .
\ee
On the other hand, at strong (but not so strong) coupling, the same applies for
\be
\mathcal F(h) \approx \frac{7}{2\pi^2} \, \zeta(3) \, h + \frac23 \, \frac1h \, , \quad h \gg 1 \, .
\ee

The solutions of \eqref{betaHLO} in these regimes corresponds to the leading order values we already reported in \eqref{leadingbeta}. Moreover,
$c_0$ can be computed in a moment in both cases. We conclude that
\begin{subequations}
\begin{align}
&g_{\tau\tau}(0) \, \beta_H^2 \approx 8 \pi^2 \alpha' \(1 + \frac{16}{\pi\sqrt{2\pi}} \frac{e^{-\frac{1}{\eta}}}{\sqrt{\eta}}\) \(1 + c_0 \, s\) \, , & \eta \ll 1 \, ,\\
&G_{\tau\tau}(0) \, \beta_H^2 \approx a^2 \ell_p^2 \(1+\frac{2}{9 a } \, \frac{2^{2/3}}{\pi^{2/3}} \, \frac{1}{\eta^{4/3}}\) (1+c_0 \, s)  \, , & \eta \gg 1 \, ,
\end{align}
\end{subequations}
where, keeping just the first subleading corrections in $\eta$,\footnote{In particular, we neglect exponentially suppressed contributions using the approximated relation $h \approx 2\pi\eta$ (see footnote \ref{strategy}).}
{\small{
\begin{subequations}
\label{c0asym}
\begin{align}
&c_0 \approx - d \(\frac12 G_2^{(A)} + G_2^{(0)}\)^{\hspace{-4pt}1/2} \[ 1 + \frac{8}{(2\pi)^{3/2}} \frac{e^{-\frac{1}{\eta}}}{\sqrt{\eta}} \(1-\frac2\eta\) \hspace{-4pt} \(1 + \frac{G_2^{(A)}-G_2^{(0)}}{G_2^{(A)} + 2 \, G_2^{(0)}}\) \]\, , \quad & \eta \ll 1 \, ,\\
&c_0 \approx - d \(\frac23 G_2^{(0)}\)^{\hspace{-4pt}1/2} \(1+\frac{1}{9 a } \, \hspace{-2pt} \frac{2^{2/3}}{\pi^{2/3}} \frac{G_2^{(A)}+G_2^{(0)}}{G_2^{(0)}} \frac{1}{\eta^{4/3}}\) \, , & \eta \gg 1 \, .
\end{align}
\end{subequations}
}}

Above, we kept the expansion in powers of $s$ explicit. However, we can also expand these results by plugging the asymptotic expressions of $s$ in \eqref{asyms} and the values of $c_0$ provided in \eqref{c0asym}. At weak coupling, our final result for $\beta_H$ up to NLO in $\sqrt{\alpha'}$ is\footnote{Given the dictionary in footnote \eqref{dictionary} and $g_{\tau\tau} = g_0^{(0)} (1 + g_2^{(0)} r^2)$, we have $\frac12 G_2^{(A)} + G_2^{(0)}=g_2^{(0)}$. Therefore, neglecting the $g_s$-corrections, we recover the result for the NLO Hagedorn temperature in Type II superstring theories (e.g., see \cite{Urbach:2023npi}), that is $g_{\tau\tau}(0) \, \beta_H^2 \approx 8 \pi^2 \alpha' \(1-\frac{d}{\sqrt{2}} \(g_2^{(0)}\)^{\hspace{-2pt}1/2} \hspace{-4pt} \sqrt{\alpha'}\) \, , \quad g_{eff} \approx 0$.}
{{
\begin{subequations}
\label{finalNLObetaH}
\begin{align}
&\beta_H^2 \approx \frac{8 \pi^2 \alpha'}{g_0^{(0)}} \(1 + \frac{32}{(2\pi)^{3/2}} \frac{e^{-\frac{1}{\eta}}}{\sqrt{\eta}}\) 	\hspace{-4pt} 
\Biggl[ 1- \sqrt{\frac{\alpha'}{2} g_2^{(0)}}\, \cdot \label{weaks}  \\
& \qquad  \qquad  \qquad  \qquad  \qquad  \qquad \qquad \qquad \cdot \[ 1 - \frac{12}{(2\pi)^{3/2}} \frac{e^{-\frac{1}{\eta}}}{\sqrt{\eta}} \frac{G_2^{(0)}}{g_2^{(0)}}  \(1 + \frac{2}{3\eta} \frac{G_2^{(A)}\hspace{-4pt}-G_2^{(0)}}{G_2^{(0)}}\) \] \Biggr] \, . \nonumber \\
\intertext{At strong coupling, our final result for $\beta_H$ up to NLO in $\ell_p^2/R_{11}$ is}
&\beta_H^2 \approx \frac{a^2 \ell_p^2}{G_0^{(0)}} \(1+\frac{2}{9 a } \, \frac{2^{2/3}}{\pi^{2/3}} \, \frac{1}{\eta^{4/3}}\) \[1- d \(\frac23 G_2^{(0)}\)^{\hspace{-4pt}1/2} \hspace{-4pt} b \, \frac{\ell_p^2}{R_{11}} \(1+\frac{1}{9 a } \, \hspace{-2pt} \frac{2^{2/3}}{\pi^{2/3}} \frac{G_2^{(A)}}{G_2^{(0)}} \frac{1}{\eta^{4/3}}\)\]  \, . \label{strongs}
\end{align}
\end{subequations}
}}

Notice that formulae \eqref{finalNLObetaH} are the corrected version of formulae \eqref{leadingbeta}. Moreover, we find full agreement with the predictions in \eqref{betaLweak}, \eqref{NLweak} and \eqref{betaLstrong}, \eqref{NLstrong}. 

We conclude that the expansion proposed in \eqref{betaansatz}, equipped with the result for $c_0$ in \eqref{runningc0}, should be robust, at least at NLO order. Moreover, let us underline how the formalism nicely encodes both the corrections in the fluctuations scale and in $g_s$ in the definition of $s$ and in the logarithmic derivative of $\mathcal F$.

As a final remark, notice that, from a semiclassical point of view, the membrane configuration has been expanded as
\be
x= X + \xi \, .
\ee
Here, $X$ stands for a classical solution of the membrane equations of motion, while $\xi$ describes the small (quantum) fluctuations around it. Focusing on the (compact) thermal direction, we have
\be
x^0 = \frac{\beta_H}{2\pi} \sigma + \mathcal O \(\sqrt{\alpha'}\) \, .
\ee
In the deep weak coupling regime, the background configuration is clearly of the same order as the fluctuations. This is a signal of the breakdown of the semiclassical approach. On the contrary, something very interesting happens in the deep strong coupling regime. Indeed, remember that the Planck scale can be expressed in terms of $g_s$ and $\alpha'$ according to \eqref{lplanck}. This means that
\be
\text{background} \sim \mathcal{O}\(\ell_p\) = \mathcal O \(g_s^{1/3} \sqrt{\alpha'}\) \quad \text{vs} \quad \mathcal O \(\sqrt{\alpha'}\) \sim \text{fluctuations}.
\ee
In other words, the background scale appears to be parametrically much bigger than the fluctuations one. Therefore, the semiclassical quantization of the winding membrane in M-theory has the potential to be fully consistent. 
On the contrary, in the weak (but not so weak) coupling regime, the perturbative corrections in $g_s$ vanish and the non-perturbative ones are not large enough to produce a hierarchy between the background and the fluctuations (indeed, $\frac{16 \,e^{-1/\eta}}{\pi\sqrt{2\pi\eta}} \lesssim 1$). 

\section{ABJM}
\label{sec:ABJM} 

As a first example of application of the formalism of section \ref{sec:11d}, let us consider the ABJM theory \cite{Aharony:2008ug}.
The eleven-dimensional metric is\footnote{In this section we work at $g_s=1$ as in \cite{Aharony:2008ug}.}
\be 
ds^2 =\frac{R^2}{4} ds^2_{AdS_4} + R^2 \left[\frac{1}{k^2} (d\varphi + k \omega)^2 +ds^2_{CP^3} \right]\,,
\ee 
where
\be 
R =l_P (2^5 \pi^2 k N)^{1/6}\,.
\ee 
Since the eleventh dimension where one reduces to ten dimensions is $\varphi$, in this case we have $G_{AA}={\rm Exp}[4\phi/3]=2^{5/3} \pi^{2/3} \lambda^{5/3} / N^{4/3}$, where $k$ is the CS level of the field theory and $\lambda=N/k$. 
$G_{\tau\tau}$ is determined by $G_{\tau\tau}=g_{\tau\tau}/\sqrt{G_{AA}}$.
Since in AdS in coordinates such that $g_{rr}=1$ one has $g_{\tau\tau} \sim 1 + r^2 + r^4/3 +...$ and $\sqrt{g} \sim const(1+5r^2/6 +...)$ (see e.g. \cite{Bigazzi:2024biz}), we have\footnote{Here, the coefficients $G_{2,4}^{(0,A)}$ are expressed in units of a proper power of $R$.}
\bea 
&& d=3\,, \qquad G^{(A)}_0 = \frac{ 2^{5/3} \pi^{2/3} \lambda^{5/3} }{ N^{4/3}}\,, \qquad G^{(0)}_0 =\frac{N^{2/3}}{2^{5/6} \pi^{1/3} \lambda^{5/6} }\,, \nonumber \\
&& G_2^{(0)} = 1\,, \qquad G_2^{(A)} =G_4^{(A)} = 0 \,, \qquad G_2 = 5/6 \,, \qquad G_4^{(0)} = 1/3 \,.
\eea 
We also have
\be 
\eta = \frac{1}{2^{1/4}\pi^{1/2}} \left( \frac{\lambda^{5/4}}{N}\right)\,.
\ee

With these data, formulae (\ref{betaLweak}), (\ref{NLweak}), or equivalently (\ref{weaks}), expressed in units of $R_{AdS}$, 
give immediately
\be 
\beta^2_H = \alpha'\,8\pi^2 \left[1 + \frac{2^3 \sqrt{2}}{\pi^{3/2}\sqrt{\eta}}e^{-\frac{1}{\eta}}  \right] - \alpha'^{3/2} 12\sqrt{2}\pi^2 \left[1 + \frac{5\sqrt{2}}{\pi^{3/2}\sqrt{\eta}}e^{-\frac{1}{\eta}} \left(1-\frac{2}{5\eta} \right)  \right] \,, \qquad \eta \ll 1\,.
\ee
The leading-in-$\eta$ terms in this result coincide with the known expressions from the analysis of the ten dimensional theory (see e.g.~\cite{Bigazzi:2023hxt}).
The non-perturbative $\eta$ corrections are the novel result in this regime.
It would be extremely interesting to test this formula by other means (Quantum Spectral Curve or similar).  

The regime where the eleven-dimensional dual description is weakly coupled is when $\eta \gg 1$.
There are no results in the literature in this regime yet.
Formulae (\ref{betaLstrong}), (\ref{NLstrong}) (or equivalently (\ref{strongs})), (\ref{NNLstrong}) and the data above give 
\bea
\beta^2_H &=& \alpha'\, 2^{5/3} \pi^{4/3} a^2 \eta^{2/3}\left[1+ \frac{2^{5/3}}{9 \pi^{2/3}a} \frac{1}{\eta^{4/3}} \right]  -  \alpha'^{3/2}\, \sqrt{3} \, 2^{7/3} \pi^{5/3} a \eta^{1/3}\left[1+ \frac{2^{5/3}}{9 \pi^{2/3}a}  \frac{1}{\eta^{4/3}} \right] + \nonumber \\
&& + \alpha'^2 \frac92 \pi^2 \left[1+ {\cal O}(\eta^{-4/3}) \right ]\,,  \qquad \eta \gg 1\,. 
\eea

\begin{figure}
	\begin{center}
		\scalebox{0.39}{
		\includegraphics{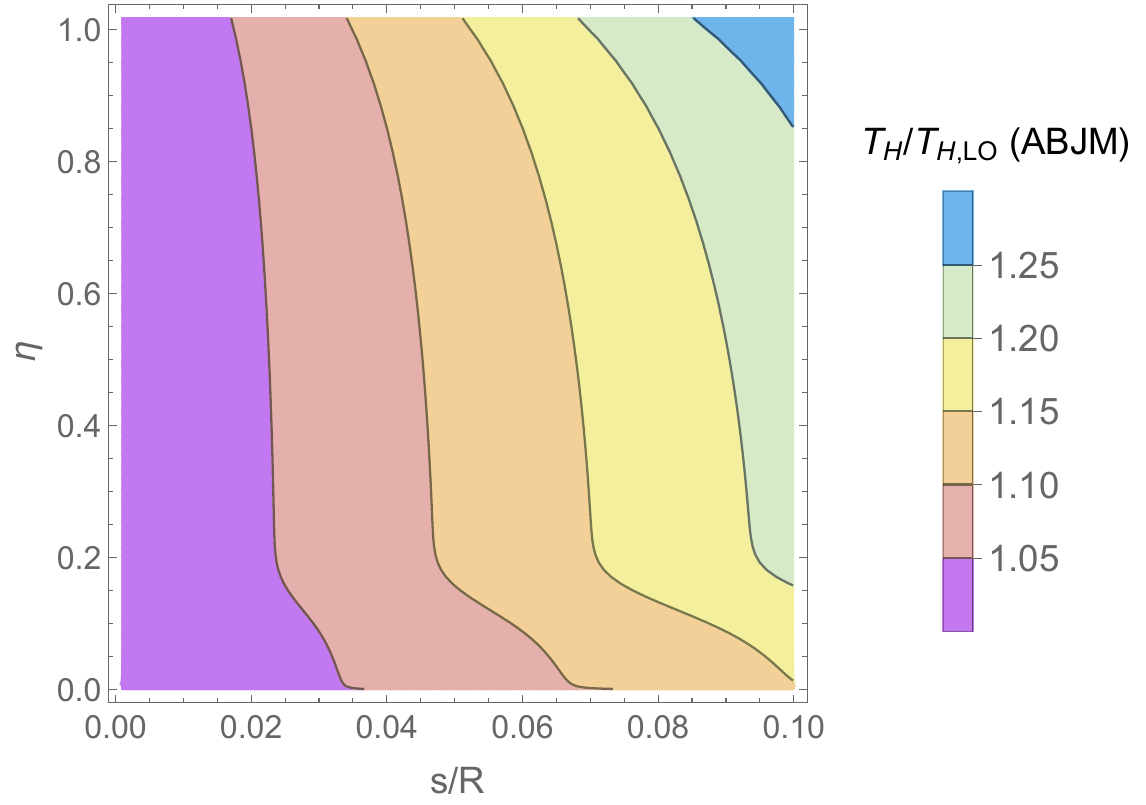}
		}
		\scalebox{0.4025}{
		\includegraphics{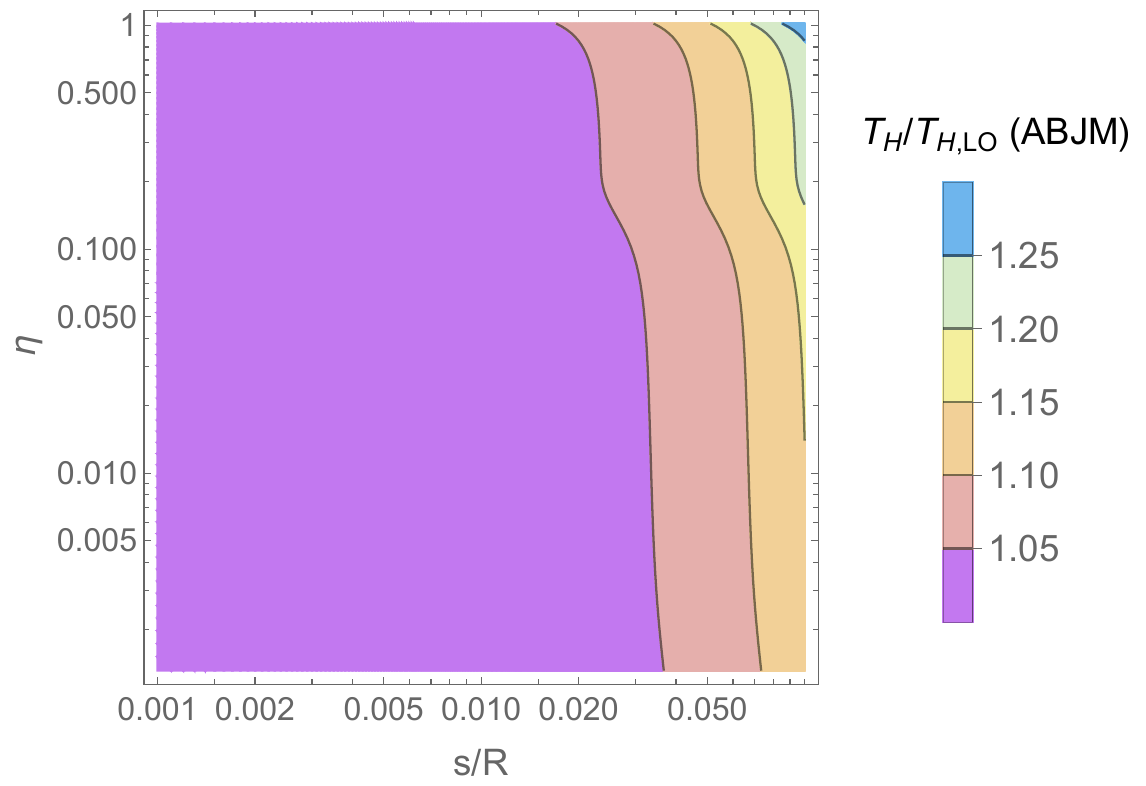}
		}
		\caption{Plots of the Hagedorn temperature in ABJM over its leading order value as a function of $s$ and $\eta$, from the NLO result \eqref{plottable} equipped with \eqref{runningc0}. There, $c_0$ depends on $h$ (see \eqref{runningepsilon}), but we can take a value of $\eta$ for any $h$ by means of \eqref{gsh} combined with \eqref{parameters}. 		 Notice that $s$ and $(G_2^{(0,A)})^{-1/2}$ are expressed in units of the same length scale, for instance the AdS radius. Finally, we truncated both the sum over $k$ and $n$ in the definition of $\mathcal{F}$ in \eqref{fullglorycalF} at $|k|$, $|2n-1|=125$. Let us stress that $s$ is the running expansion parameter and it is supposed to be small at any scale $g_s$. The latter, instead, can flow from the weak to the strong coupling regime.}
		\label{fig:betaHABJM}
	\end{center}
\end{figure}

For intermediate values of $\eta$, the leading-$\alpha'$-order result for $T_H$ is as in figure \ref{fig:flatall}, trading the dependence on $g_s$ with that on $\eta$. 
We can use the procedure of section \ref{sec:runningparameter} to study numerically $T_H$ in this regime as a function of both $\eta$ and $s$ (the expansion parameter defined in (\ref{s})).
As an example, in figure \ref{fig:betaHABJM} we report the Hagedorn temperature over its leading order value as a function of $s$ and $\eta$ 
in the ABJM case.

\section{WYM and finite $\theta$-angle effects}
\label{sec:WYM}

In this section we study the Hagedorn temperature of the non-supersymmetric, confining WYM theory \cite{Witten:1998zw}, including the dependence on a non-trivial $\theta$-angle. 
We work in the eleven-dimensional setup, which includes the ten-dimensional results at weak coupling.
The eleven-dimensional metric dual to the WYM theory is basically an AdS soliton.
We use the notations in \cite{Dubovsky:2011tu} and \cite{Bigazzi:2015bna}, but we work with dimensionful coordinates, such that
\be \label{metricwym}
ds^2 = \alpha' (\pi g_s N)^{2/3} \left[ \frac{2u}{\alpha' \beta_4 \lambda} \left(dx_{\mu}dx^{\mu} + \frac{f}{H_0} d\chi^2 \right) +\frac{du^2}{f u^2} + d\Omega_4^2\right] + \frac{\beta_4 \lambda u H_0}{2\alpha' (\pi g_s N)^{4/3}} (dz+A d\chi)^2\,,
\ee 
where $z$ is the eleventh direction with periodicity $2\pi R_{11}$, $\chi$ has periodicity $2\pi \beta_4 \equiv 2\pi M_{KK}^{-1}$, $A$ is some function of $u$ we are not going to need and
\be 
\beta_4 = \frac{2\alpha' \lambda}{9 u_0}\frac{1}{1+\Theta^2}\,, \quad \lambda = g_{YM}^2 N = 2\pi g_s N l_s M_{KK}\,, \quad f = 1-\frac{u_0^3}{u^3}\,, \quad H_0 =1-\frac{u_0^3}{u^3}\frac{\Theta^2}{1+\Theta^2}\,.
\ee
In these expressions 
\be 
\Theta = \frac{\lambda}{4\pi^2} \left(\frac{\theta+2\pi k}{N} \right)\,, \qquad k \in \mathbb{Z}\,,
\ee
contains the dependence on the $\theta$-angle in the planar limit, $k$ accounts for the various branches of vacua and a minimization over $k$ of the vacuum energy at fixed $\theta$ is understood.
For $\Theta=0$ the results for trivial $\theta$-angle are recovered.

We thus have
\be 
G_{AA} = \frac{\beta_4 \lambda u H_0}{2\alpha' (\pi g_s N)^{4/3}} = e^{4\phi/3} = \frac{u}{R}H_0\,, \qquad G_{\tau\tau} =  \frac{2u(\pi g_s N)^{2/3}}{\beta_4 \lambda} = \frac{g_{\tau\tau}}{\sqrt{G_{AA}}}\,, 
\ee
with
\be 
R= l_s(\pi g_s N)^{1/3}\,.
\ee 
We pass to a radial variable $r$ such that in 10d $g_{rr}=1$ by the change 
\be 
u-u_0 = \frac34 m_0 r^2\sqrt{1+\Theta^2}\left(1-\frac{r^2\sqrt{1+\Theta^2}(1+3\Theta^2)}{8\sqrt{u_0 R^3}} \right)\,, \qquad m_0 = \frac23 M_{KK} = \sqrt{\frac{u_0(1+\Theta^2)}{R^3}}\,.
\ee 
Thus the data we need are 
\bea 
&& G_0^{(A)} =\frac{u_0}{R(1+\Theta^2)}\,, \quad G_0^{(0)}=\frac{u_0}{R}\,, \quad d=2 \,, \\ && G_2^{(A)}=\frac{3(1+3\Theta^2)\sqrt{1+\Theta^2}}{4R^\frac{3}{2}\sqrt{u_0}} \,,  
\quad G_2^{(0)}=\frac{3\sqrt{1+\Theta^2}}{4R^\frac{3}{2}\sqrt{u_0}}\,, \quad G_2 = \frac{(5-3\Theta^2)\sqrt{1+\Theta^2}}{4\sqrt{u_0}R^\frac{3}{2}}\,, \nonumber \\
&& G_4^{(A)}= -\frac{3}{32R^3 u_0}(1+\Theta^2)(1+24 \Theta^2 + 9 \Theta^4)\,, \quad G_4^{(0)} = -\frac{3}{32 R^3u_0}(1+4 \Theta^2 + 3 \Theta^4)\,.\nonumber
\eea

We can plug these data in formulae  (\ref{betaLweak}), (\ref{NLweak}) or (\ref{weaks}) to get the weak coupling value of the inverse Hagedorn temperature up to NLO in the $\alpha'$ expansion, which can be written as
\begin{eqnarray} \label{wymweak}
\beta^2_H = \frac{4\pi}{T_s} \left[ \left(1 + \frac{2^{7/2} }{\pi^{3/2}\sqrt{\eta}}e^{-\frac{1}{\eta}} \right) - \frac{M_{KK}}{\sqrt{2\pi T_s}}  \left(1 +  \frac{2^{3/2} }{\pi^{3/2}\sqrt{\eta}}\frac{3+4\Theta^2}{1+\Theta^2} e^{-\frac{1}{\eta}} \left(1-\frac{2+4\Theta^2}{\eta(3+4\Theta^2)} \right) \right)  \right]  && \nonumber \\
&& \hspace{-2.5cm}\eta \ll 1\,,
\end{eqnarray}
where for the string tension we have
\be 
T_s = \frac{1}{2\pi \alpha'}\left( \frac{u_0}{R}\right)^{3/2}\frac{1}{\sqrt{1+\Theta^2}} = \frac{2}{27\pi}\lambda M_{KK}^2 \frac{1}{(1+\Theta^2)^2}\,, 
\ee
and in this case
\be 
\eta = \frac{1}{6^{3/2}\pi^{2}}\frac{\lambda^{3/2}}{N}\frac{1}{(1+\Theta^2)^{3/2}}\,.
\ee
The result (\ref{wymweak}) reproduces the known NLO result at leading order in $\eta$ for $\theta=0$ \cite{Urbach:2023npi}, but extends that result, which can be derived directly from the string in ten dimensions, at finite $\theta$ and includes the first non-perturbative corrections in $\eta$, corresponding to instantonic contributions.
In principle one could go on and calculate the NNLO corrections.
For example at leading order in $\eta$, working in ten dimensions and making use of the formulae in \cite{Bigazzi:2024biz} one gets
\be \label{10dresult}
\beta^2_H = \frac{4\pi}{T_s} \left[ 1  - \frac{M_{KK}}{\sqrt{2\pi T_s}} -\frac14 \left(\frac{M_{KK}}{\sqrt{2\pi T_s}} \right)^2 \frac{1-8\log{2}-\Theta^2(3+8\log{2})}{1+\Theta^2}  \right] \,, \qquad \eta \ll 1\,.
\ee

At strong coupling, formulae (\ref{betaLstrong}), (\ref{NLstrong}) (or (\ref{strongs})) and (\ref{NNLstrong}) give
\bea\label{wymstrong}
\beta^2_H & = & \frac{4\pi}{T_s}  \frac{a^2}{2^{4/3} \pi^{2/3}} \eta^{2/3} \Biggl[ \left(1+ \frac{ 2^{5/3} }{9a \pi^{2/3}} \frac{1}{\eta^{4/3}} \right) + \\
&& -  \frac{2^{5/3} \pi^{1/3}}{3 a \sqrt{1+\Theta^2}} \left(\frac{M_{KK}}{\eta^{1/3}\sqrt{2\pi T_s}}\right) 
 \left(1+ \frac{ 2^{2/3}}{3a  \pi^{2/3}} \frac{(1+\Theta^2)}{\eta^{4/3}} \right) + \nonumber\\
 && - \frac{2^{1/3} \pi^{2/3}}{3^4 a^2 \sqrt{1+\Theta^2}} \left(\frac{M_{KK}}{\eta^{1/3}\sqrt{2\pi T_s}}\right)^2 (23-96\Theta^2-36\Theta^4) \left(1+ {\cal O}(\eta^{-4/3}) \right) \Biggr] \,,  \quad \eta \gg 1\,. \nonumber
\eea
Let us stress that in the third line we reported just the contribution of the effective field theory.

In the intermediate regime of $\eta$, the LO result for $T_H/\sqrt{2\pi T_s}$ is the one reported in figure \ref{fig:flatall}, trading the dependence on $g_s$ with that on $\eta$. 
At any fixed value of $\eta$ the NLO corrections can be calculated with the formalism of section \ref{sec:runningparameter}.

Formulae (\ref{wymweak}), (\ref{wymstrong}) are written in the two ``natural'' expansion parameters at $\theta=0$, i.e.~$\eta$ and $M_{KK}/\sqrt{2\pi T_s}$.
As one can see, while the $\theta$-dependence enters $\beta_H$ at LO at large $\eta$ (remember that $\eta$ depends on $\theta$), at small $\eta$ it enters at LO only including the first non-perturbative correction, otherwise it enters at NNLO (see (\ref{10dresult})).
Nevertheless, as usual the results of the comparison of different theories, in this case theories at different values of $\theta$, depend crucially on the adopted comparison scheme.
The one above, where $M_{KK}/\sqrt{T_s}$ is kept fixed, is only one of the possibilities.

As a relevant example, one can write
\be 
\frac{M_{KK}}{\sqrt{T_s}} = \sqrt{\frac{27\pi}{2}} \frac{(1+\Theta^2)}{\sqrt{\lambda}}\,,
\ee
and keep fixed the UV coupling $\lambda$ instead.
In this case $M_{KK}/\sqrt{T_s}$ depends explicitly on $\theta$, which thus enters the Hagedorn temperature already at NLO even for small $\eta$, as
\bea 
&& \beta^2_H = \frac{4\pi}{T_s} \left[ 1  - \frac{3^{3/2}}{2} \frac{(1+\Theta^2)}{\sqrt{\lambda}}  \right]\,, \qquad \qquad \qquad  \qquad \qquad \ \ \eta \ll 1\,,\\
&& \beta^2_H  =  \frac{4\pi}{T_s}  \frac{a^2}{2^{4/3} \pi^{2/3}} \eta^{2/3} \left[ 1  -  \frac{2^{2/3} 3^{1/2} \pi^{1/3}}{ a \eta^{1/3}} \frac{\sqrt{1+\Theta^2}}{\sqrt{\lambda}}  \right]\,, \qquad  \eta \gg 1\,. 
\eea 
Thus, we conclude that at fixed UV coupling $\lambda$ (and fixed $N$), the Hagedorn temperature $T_H/\sqrt{T_s}$ \emph{increases} with $\theta$ both at small (at NLO) and large (at LO) $\eta$.\footnote{Another sensible choice of comparison scheme is the one where the ratio $M_{gb}/\sqrt{T_s}$ is kept fixed, where $M_{gb}$ is the mass of the lightest glueball, whose behavior with $\theta$ is (see e.g. \cite{Bigazzi:2015bna})
$M_{gb}(\Theta) = M_{gb}(\Theta=0)/\sqrt{1+\Theta^2}$.
}

We can also compare the Hagedorn temperature with the critical temperature for deconfinement. 
Upon double Wick rotation in the $t$ and $\chi$ coordinates, the metric in (\ref{metricwym}) is translated into a black brane solution.
At $\theta=0$ the geometry of the two coordinates is exactly interchanged by the rotation.
Since the temporal circle has length $\beta$ and the $\chi$ circle has length $2\pi \beta_4$, there is a Hawking-Page transition between the two metrics at the critical temperature $T_c = 1/2\pi \beta_4 = M_{KK}/2\pi$.
This value does not depend on the regime of $\eta$ and in fact it is the one encountered in ten dimensions.\footnote{At $\theta \neq 0$ the critical temperature in the small $\eta$ limit depends on $\Theta$ \cite{Bigazzi:2015bna}, and $T_c/\sqrt{2\pi T_s} = 3^{3/2} \sqrt{1+\Theta^2} / 4\pi \sqrt{\lambda}$, which increases with $\theta$ at fixed $\lambda$. Note that this is a leading order effect, as opposed to the analogous subleading effect on $T_H$ described before.}
We can thus compare 
\be \label{Tcwym}
\frac{T_c}{\sqrt{2\pi T_s}} = \frac{3^{3/2}}{4 \pi}\frac{1}{\sqrt{\lambda}}\,,
\ee
with $T_H/\sqrt{2\pi T_s}$ of figure \ref{fig:flatall}.
\begin{figure}
		\center
\scalebox{0.8}{\includegraphics{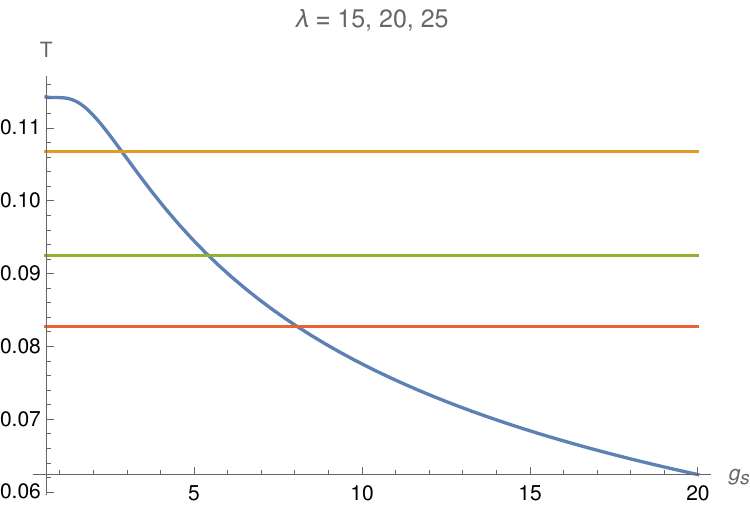}}
\caption{The Hagedorn temperature $T_H$ (blue curve) versus the critical temperature $T_c$  for different values of $\lambda$ (constant functions, from top to bottom $\lambda = 15, 20, 25$) as a function of $g_s=2^{3/2}\pi \eta$. Both are normalized with the string tension.}
		\label{fig:intersectionh}
\end{figure} 
The result is shown in figure \ref{fig:intersectionh}. 	
While $T_H$ is parametrically (in $\lambda$) larger than $T_c$ at small $\eta$, eventually by increasing $\eta$ the two values will coincide.
The intersection point depends on $\lambda$.
In general, the Hagedorn temperature is the limiting temperature of the metastable branch of the low-temperature phase (graviton gas phase) above the Hawking-Page temperature $T_c$.
So, considering the regime of parameters where $T_H < T_c$ does not make sense: the intersection point must be viewed as the limiting parameter values at which there exists a Hagedorn temperature.

As a divertissement, let us consider the intersection point in ``real world'' YM, pretending that $N=3 \sim \infty$.
There we have from the lattice that, around $T_c$, $g_{YM}^2/4\pi = \alpha_s \sim 1/2$ \cite{Borsanyi:2013bia}, thus $\lambda = 6\pi$ and $\eta =1/3 \sqrt{\pi}$. 
Our numerical curve for $T_H$ and formula (\ref{Tcwym}) for $T_c$ would give for these numerical values roughly\footnote{We use the numerics corresponding to the curve for $T_H$ as a function of the variable $g_s$. This is the flat space curve. In a generic curved background, at leading order the curve is numerically the same, but the variable is $2^{3/2}\pi \eta$ rather than $g_s$, due to the rescaling of $g_{eff}$ in (\ref{prescri}) (and using the definition of $\eta$ in (\ref{eta})). So, on curved backgrounds we can read the value of $T_H$ from the flat space curve just by trading $g_s$ for $2^{3/2}\pi \eta$. In the present case of the WYM theory and for $\eta = 1/3 \sqrt{\pi}$, we have to read the value of $T_H$ from the curve at $g_s = 2^{3/2}\pi \eta \sim 1.67$ (of course in this expression $g_s$ is just a symbol, it is not the actual value of the string coupling). For this value the curve of $T_H$ is basically equal to the $g_s=0$ value $1/2^{3/2}\pi$, which is also attained at large $N$ with fixed $\lambda$.} $T_H/\sqrt{2\pi T_s} \sim 0.113$, $T_c/\sqrt{2\pi T_s} \sim 0.095$, which translates into $T_H/T_c \sim 1.19$.
This value is not far from the value reported by lattice studies for  pure YM.
In fact, the latter is roughly $T_H/T_c \sim 1.1$ in the large $N$ limit \cite{Bringoltz:2005xx}.
Moreover, using the observation in \cite{Caselle:2015tza} that for $SU(3)$ YM on the lattice it seems that $T_H$ is well approximated by the bosonic string value, $T_H/\sqrt{T_s} \sim \sqrt{3/2\pi}$, and employing the known result for the critical temperature of the same theory, $T_c/\sqrt{T_s} \sim 0.63$ (see e.g.~\cite{Boyd:1996bx}), one obtains once again $T_H/T_c \sim 1.1$.

\section{Flavor effects on the Hagedorn temperature}
\label{sec:flavors}

In this section the dependence of the Hagedorn temperature $T_H$ on the flavor contribution is derived in confining holographic models.
We work in ten dimensions and consider in turn the examples of the backreacted Witten-Sakai-Sugimoto (WSS), Maldacena-Nu\~nez (MN) 
and Klebanov-Strassler (KS) theories. 
These are (almost, see footnote \ref{footno}) all the available models with dynamical flavors which are regular in the dual IR region, where the thermal scalar mode is localized.
In all the cases the flavors reduce the value of $T_H/\sqrt{T_s}$ at fixed coupling.
For the WSS theory the deconfinement temperature $T_c$ is known, so we can compare the flavor effects on the two temperatures. 
When the flavors are massive, one can consistently reduce their effects by increasing the mass.

In this section the Hagedorn temperature is expressed according to formula (1.4) in \cite{Bigazzi:2023hxt}, which at NLO reads
\begin{equation}\label{NLO}
	\frac{T_s}{2}\beta^2_H =2\pi \left( 1- \frac{d}{2} \mu \right)\,,
\end{equation}
where $T_s$ is the string tension, $\beta_H=1/T_H$, $d=\{2 \ (\text{WSS}),\ 3 \ (\text{MN, KS})\}$, and 
\begin{equation}\label{muhere}
	\mu = \frac{\beta_H}{2\pi}\frac{\sqrt{2\pi \alpha' T_s}}{l}\,.
\end{equation}
The parameter $l$ can be read from the metric expressed in the form (1.1) in \cite{Bigazzi:2023hxt}
\begin{equation}\label{metric}
	ds^2 = 2\pi \alpha'T_s \left(1+\frac{r^2}{l^2} \right) \left(-dt^2 +dx_idx^i \right)+ dr^2 + ... \,.
\end{equation}
We are going to compare the Hagedorn temperature in the unflavored theory with the one in the flavored theory by expressing them in terms of the respective string tensions.
Solving (\ref{NLO}) we get
\begin{equation}\label{hagedorn}
	\frac{T_H}{\sqrt{T_s}} = \frac{1}{2\sqrt{\pi}} + \frac{d\sqrt{\alpha'}}{4 l \sqrt{2\pi}}\,.
\end{equation}
The qualitative behavior of the flavor corrections are then contained in the qualitative behavior of the parameter $l$.

\subsection{Witten-Sakai-Sugimoto model}

The gravity solution dual to the WSS model with backreacting (smeared) flavors can be found in \cite{Bigazzi:2014qsa}.\footnote{The non-smeared solution in \cite{Burrington:2007qd} is not suitable for our computation since it does not have a regular IR dual to the confining regime.}
The solution contains the flavor corrections up to first order in the parameter
\begin{equation}\label{epsilon}
	\epsilon_f = \frac{1}{12\pi^3} \lambda^2 \frac{N_f}{N} \,,
\end{equation}
where $\lambda, N_f, N$ are the 't Hooft coupling, the number of flavors and colors and it is assumed that $\epsilon_f \ll 1$.
The string frame metric is presented in \cite{Bigazzi:2014qsa} in the form 
\begin{equation}
	ds^2 = e^{2\hat\lambda} \left(-dt^2 +dx_idx^i \right)+ l_s^2 e^{-2\varphi}d\rho^2 + ... \,,
\end{equation}
where $\varphi=2\phi - 4 \hat\lambda - \tilde\lambda - 4\nu$ and all these functions depend only on the radial variable $\rho$.  
Their form in the IR region is
\begin{eqnarray}\label{solIR}
	&& \hat \lambda = f_0(\tilde r) + \frac34 \log{\frac{u_0}{R}} +  \frac{\epsilon_f }{4} \left(3A_1-4k \right) +... \,, \nonumber \\
	&& \tilde \lambda = f_0(\tilde r) + \frac34 \log{\frac{u_0}{R}} - \frac32 \tilde r  - \frac{\epsilon_f }{4} \left(A_1+4k \right)+...  \,, \nonumber \\
	&& \phi = f_0(\tilde r) + \frac34 \log{\frac{u_0}{R}} + \log{g_s} +   \frac{\epsilon_f}{4} \left(11A_1-4k \right) +...  \,,  \nonumber \\
	&& \nu = \frac13 f_0(\tilde r) + \frac14 \log{\frac{u_0}{R}} + \log{\frac{R}{l_s}} + \frac{\epsilon_f}{12} \left(11A_1-20k \right) +...  \,, 
\end{eqnarray}
where
\begin{equation}
	f_0(\tilde r) = -\frac{1}{4} \log{\left(1-e^{-3\tilde r} \right)}  \,,  \qquad \tilde r = \frac{u_0^3}{l_s^3 g_s^2} \rho \,.
\end{equation}
In (\ref{solIR}) we have already selected the values of some constants of integration which correspond to the less-singular solution.
Moreover, we have
\begin{eqnarray}
	&& A_1 = \frac{81 \sqrt{3} \pi^2 (-9+\sqrt{3} \pi -12 \log{2} + 9\log{3}) }{43120 \times 2^{2/3} \Gamma[-14/3]\Gamma[-2/3]^2} \,, \nonumber \\
	&& k = \frac{\pi^{3/2} (3+\sqrt{3} \pi -12 \log{2} + 9\log{3}) }{78  \Gamma[1/6]\Gamma[-2/3]} \,.
\end{eqnarray}

In order to pass the metric to the form (\ref{metric}) the relation between $r$ and $\rho$ in the IR (which corresponds to the large $\rho$ regime) must be derived from $dr=l_s e^{-\varphi}d\rho$, with the result 
\begin{equation}
	r = \frac23 R^{3/4}u_0^{1/4} e^{ -\frac{3u_0^3}{2 g_s^2 l_s^3} \rho } \left[1+ \epsilon_f\left(\frac{11}{12}A_1 -\frac{29}{3} k \right) \right]\,.
\end{equation}
In this coordinate the $g_{tt}$ component of the metric reads
\begin{equation}
	g_{tt}(r) = \left(\frac{u_0}{R} \right)^{3/2} \left[1+ \epsilon_f\left(\frac{3}{2}A_1 -2 k \right) \right] \left( 1+ \frac{9\left[1+ \epsilon_f\left(-\frac{11}{6}A_1 +\frac{58}{3} k \right) \right]r^2}{8 R^{3/2}u_0^{1/2}}  \right) \,,
\end{equation}
such that comparing with (\ref{metric}) we get
\begin{equation}\label{tension}
	2\pi\alpha' T_s = \left(\frac{u_0}{R} \right)^{3/2} \left[1+ \epsilon_f\left(\frac{3}{2}A_1 -2 k \right) \right]\,, \qquad l= \frac23 \sqrt{2}R^{3/4}u_0^{1/4}  \left[1+ \epsilon_f\left(\frac{11}{12}A_1 -\frac{29}{3} k \right) \right] \,.
\end{equation}

We are going to compare the Hagedorn temperature in the unflavored theory with the one in the flavored theory by expressing them in terms of the respective string tensions, which are given by (\ref{tension}) with and without imposing $\epsilon_f=0$.
Looking at formula (\ref{NLO}), this implies that at leading order (i.e.~posing $\mu=0$) the two ratios of $T_H / \sqrt{T_s}$ are equal.
Instead, by solving  (\ref{NLO}) at NLO and using (\ref{tension}) we find
\begin{equation}\label{hage}
	\frac{T_H}{\sqrt{T_s}} = \frac{1}{2\sqrt{\pi}} + \frac{3\sqrt{3}}{8 \sqrt{\pi \lambda}}  \left[1+ \epsilon_f\left(-\frac{11}{12}A_1 +\frac{29}{3} k \right) \right] \sim \frac{1}{2\sqrt{\pi}} + \frac{3\sqrt{3}}{8 \sqrt{\pi \lambda}}  \left(1-0.35 \epsilon_f \right)\,.
\end{equation}
Thus, the ratio of the Hagedorn temperature with the square root of the string tension \emph{decreases} in the flavored theory if we keep fixed the 't Hooft coupling.

As a comparison, the critical temperature for deconfinement $T_c$ from \cite{Bigazzi:2014qsa} reads
\begin{equation}
	T_c = \frac{M_{KK}}{2\pi} \left[1+ 1.987 \frac{1}{126\pi^3} \lambda^2 \frac{N_f}{N}  \right] \,.
\end{equation}
By normalizing it with the square root of the string tension we find
\begin{equation}
	\frac{T_c}{\sqrt{T_s}}  \sim \frac{3\sqrt{3}}{2 \sqrt{2 \pi \lambda}}  \left(1-0.38 \epsilon_f \right)\,.
\end{equation}
One can observe that this ratio has a very similar form to the NLO term in (\ref{hage}), and that the critical temperature for deconfinement decreases slightly faster than the Hagedorn temperature: the flavors seem to push the two temperatures apart, specifically
\be 
\frac{T_H}{T_c} \sim \sqrt{\frac{2\lambda}{27}} \left[ 1 + \frac{3\sqrt{3}}{4 \sqrt{\lambda}} + \epsilon_f \left(0.38 +0.03 \frac{3\sqrt{3}}{4 \sqrt{\lambda}}  \right) \right]\,.
\ee 

An important comment is in order.
As usual, when comparing theories with different number of flavors, a crucial point is the choice of physical quantities which are kept to be the same in the two theories.
Typically, the answers will depend even qualitatively from this choice.
In this sense, the form of expression (\ref{hage}) is tailor-made for a comparison where the quantity to be kept fixed is $\lambda$.
This is a very natural parameter to keep fixed when varying $N_f$, since the parameter $\epsilon_f$ (\ref{epsilon}) depends on it as well. 
But other choices are legitimate too.
For example, in the unflavored case the result for the NLO term of $T_H$ can be naturally expressed in terms of the ratio $M_{KK}/\sqrt{2 \pi T_s}$. 
In the flavored case this ratio reads (see formula (3.40) of \cite{Bigazzi:2014qsa})
\be 
\frac{M_{KK}}{\sqrt{2\pi T_s}} = \sqrt{\frac{27}{4 \lambda}}\left[1-\frac{\epsilon_f}{2} (3A_1-A_2-28k)  \right] \sim \sqrt{\frac{27}{4 \lambda}}\left[1-0.57 \epsilon_f \right]\,,
\ee 
i.e.~at fixed value of the coupling the ratio decreases.
In terms of this ratio one could write
\begin{equation}
	\frac{T_H}{\sqrt{T_s}} = \frac{1}{2\sqrt{\pi}} + 0.14 \frac{M_{KK}}{\sqrt{2\pi T_s}} [1+0.05\epsilon_f ] \,.
\end{equation}
In this way the correction in the number of flavors increases $T_H/\sqrt{T_s}$ if $M_{KK}/\sqrt{2 \pi T_s}$ is kept fixed.

\subsection{Flavored Maldacena-Nu\~nez theory}

The same computation done in the WSS model can be performed in the massive-flavored MN solution in  \cite{Conde:2011rg} 
(the massless solution is singular in the IR).

The functions appearing in the metric have the following asymptotic in the IR  ($\rho \sim 0$)\footnote{The notation in \cite{Conde:2011rg} is such that $g_s = \alpha'=1$. We take accordingly $e^{\phi_0}=1$.}
\begin{eqnarray}
	&& e^{2k} = N \left( \beta_f + \frac{4}{5\beta_f}(\beta_f^2-1)\rho^2 + ... \right)\,, \nonumber \\
	&& e^{4\phi} = \frac{4 }{N^3\beta_f^3}\left(1+\frac{16}{9\beta_f^2}\rho^2 + ... \right) \,.
\end{eqnarray}
In these expressions the parameter $\beta_f$ (called simply $\beta$ in \cite{Conde:2011rg}) carries the information about the flavors, depending explicitly on $N_f/N$.
The relevant part of the metric in string frame reads
\begin{equation}
	ds^2 = e^{\phi} \left(-dt^2 +dx_idx^i \right)+  e^{\phi +2k}d\rho^2 + ... \,.
\end{equation}
Again, we change coordinate from $\rho$ to $r$ in the IR, obtaining
\begin{equation}
	r = (4 \beta_f N )^{1/8} \rho + ... \,,
\end{equation}
such that the metric is written as in (\ref{metric}) with
\begin{equation}
	l = \frac{3 N^{1/8}}{2^{3/4}} \beta_f^{9/8}\,.
\end{equation}
From figure 3 of \cite{Conde:2011rg} we learn that $\beta_f \geq 1$, with the value $\beta_f=1$ corresponding to the unflavored theory.
As stressed in \cite{Conde:2011rg}, $\beta_f$ is the natural parameter to measure the effects of flavors in the IR of this theory. 
Thus, $l$ is an increasing function of the number of flavors, such that the Hagedorn temperature is decreased by the flavors, as in the WSS case.
Moreover, from figure 3 of \cite{Conde:2011rg} one can clearly see that when the mass of the flavors is increased, $\beta_f$ is reduced: the more the flavors are massive, the less important is their effect on $T_H$. 

\subsection{Flavored Klebanov-Strassler theory}

\begin{figure}[ht!]
\centering
\includegraphics[width=0.8\linewidth]{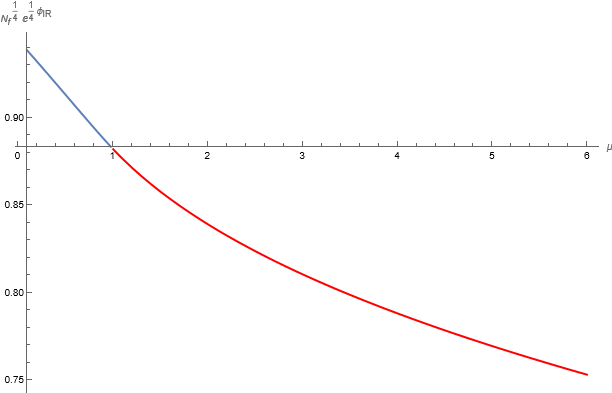}
   \caption{$N_f^\frac{1}{4}e^{\frac{\phi_{IR}}{4}}$ as a function of $\mu$.}
\label{fig: dynamics_linear_quiver_1}
\end{figure}

The Klebanov-Strassler cascading theory is a $\mathcal{N}=1$ theory dual to D3-branes placed on the deformed conifold \cite{Klebanov:2000hb}. The addition of backreacted flavors with D7-branes was considered in \cite{Bigazzi:2008qq}. The relevant part of the string frame metric of the backreacted flavored solution is given by
\begin{eqnarray}
    ds^2=e^{\frac{\phi}{2}}\left(h(\tau)^{-\frac{1}{2}}dx_\mu dx^\mu+h(\tau)^\frac{1}{2}\frac{1}{9}e^{2G_3(\tau)}d\tau^2\right)+...
\end{eqnarray}
From \cite{Bigazzi:2008qq}, we have (in $g_s=1$ units)
\begin{eqnarray}
&e^\phi=e^{\phi_{IR}},\nonumber \\
 & h(\tau)=\frac{2^{\frac{2}{3}}\alpha'^2M^2}{\epsilon^{\frac{8}{3}}_{IR}} \bigg[h_0-e^{\phi_{IR}}\int_0^\tau d\xi \frac{(\xi \coth \xi-1)(\sinh 2\xi-2\xi)^\frac{1}{3}}{\sinh^2\xi}\bigg],  \nonumber \\
 &e^{2G_3(\tau)}=\frac{3}{2}\frac{\epsilon_{IR}^\frac{4}{3}}{K(\tau)^2}\,,
\end{eqnarray}
where 
\begin{eqnarray}
    K(\tau)=\frac{(\sinh (2\tau)-2\tau)^\frac{1}{3}}{2^\frac{1}{3}\sinh (\tau)}\,,
\end{eqnarray}
and $M$, the number of fractional branes, corresponds to the rank of the gauge group in the IR.
Expanding close to $\tau=0$ the integral appearing in the definition of $h(\tau)$, computing the integral and expanding the complete result, we get
\begin{eqnarray}
e^\frac{\phi}{2}h^{-\frac{1}{2}}\approx \frac{e^{\frac{\phi_{IR}}{2}}\epsilon_{IR}^{\frac{4}{3}}}{2^{\frac{1}{3}}\sqrt{h_0}M\alpha'}+\frac{e^{\frac{3}{2}\phi_{IR}}\epsilon_{IR}^{\frac{4}{3}}\tau^2}{6\cdot 2^\frac{2}{3}3^{\frac{1}{3}}h_0^\frac{3}{2}M\alpha'}\,,
\end{eqnarray}
and
\begin{eqnarray}
    e^\frac{\phi_{IR}}{2}h^\frac{1}{2}\frac{1}{9}\frac{3}{2}\epsilon_{IR}^\frac{4}{3}\bigg(\frac{(\sinh 2\tau-2\tau)^\frac{1}{3}}{2^{\frac{1}{3}}\sinh \tau}\bigg)^{-2}\approx \frac{e^{\frac{\phi_{IR}}{2}}\sqrt{h_0}M\alpha'}{2\cdot 6^{\frac{1}{3}}}+\mathcal{O}(\tau)^2\,.
\end{eqnarray}
From this we can define a new holographic coordinate
\begin{eqnarray}
r=\tau \frac{e^{\frac{\phi_{IR}}{4}}h_0^\frac{1}{4}M^\frac{1}{2}\alpha'^{\frac{1}{2}}}{2^\frac{2}{3} 3^\frac{1}{6}}\,,
\end{eqnarray}
and bring the metric to the form in (\ref{metric}), from which we identify
\begin{eqnarray}
    T_s=\frac{e^{\frac{\phi_{IR}}{2}}\epsilon_{IR}^{\frac{4}{3}}}{2^\frac{4}{3}h_0^\frac{1}{2}M\pi \alpha'^2}\,,
\end{eqnarray}
and
\begin{eqnarray}
    l=\sqrt{3}e^{-\frac{\phi_{IR}}{4}}h_0^\frac{3}{4}\sqrt{M\alpha'}\,,
\end{eqnarray}
from which we get
\begin{eqnarray}
\frac{T_H}{\sqrt{T_s}}=\frac{1}{2\sqrt{\pi}}+\frac{\sqrt{3}e^{\frac{\phi_{IR}}{4}}}{\sqrt{2^5\pi M}h_0^\frac{3}{4}}\,.
\end{eqnarray}
This expression depends heavily on the value of the dilaton in the IR, which is computed as an integral of the flavor profile $N_f(\tau)$ as  
\begin{eqnarray}
    e^{-\phi_{IR}}=\frac{1}{4\pi}\int_{\tau_q}^{\tau_0}N_f(\xi)d\xi\,.
\end{eqnarray}
But $N_f(\tau)=N_f \hat{f}(\tau) $ where $\hat{f}$ is a function independent of $N_f$, so $e^{\phi_{IR}}$ is inversely proportional to $N_f$. The details on the function $N_f(\xi)$ can be found in \cite{Bigazzi:2008qq}. For our purposes, it suffices to say that we are considering massive embeddings of the D7-brane, characterized by a parameter $\mu$ which\footnote{This parameter is not to be confused with the $\mu$ defined in equation (\ref{muhere}).} determines the minimum distance in the radial direction up to which the branes extend. These embeddings make up a macroscopic flavor profile which takes a different form depending on if $\mu>1$ or $\mu<1$. The explicit form of $\phi_{IR}$ when $\mu>1$ is given by
{\small
\begin{eqnarray}
&e^{-\phi^{\mu>1}_{IR}}=\frac{N_f}{4\pi}\int_{\mathrm{arcosh}(2\mu^2-1)}^{\tau_0}d\tau \bigg[1-2\mu^2\frac{2\cosh\tau}{\cosh2\tau-1}+\frac{1}{2\pi\mu^2}\bigg(8\mu^4\frac{2\cosh\tau}{\cosh2\tau-1}\arctan\sqrt{\frac{(1+2\mu^2-\cosh\tau)\coth^2\frac{\tau}{2}}{1-2\mu^2+\cosh\tau}}\nonumber \\&+2\mu^2\sqrt{4\mu^2(\cosh\tau-\mu^2)\mathrm{csch}^2\tau-1}+4i(\mu^2-1)\bigg(F[\arcsin\big(\frac{\mu\coth^2\frac{\tau}{2}}{1+\mu^2}\big),1-\frac{1}{\mu^4}]-K\big(1-\frac{1}{\mu^4}\big)\bigg)\nonumber \\&-8i\bigg(\Pi\big(1+\frac{1}{\mu^2},1-\frac{1}{\mu^4}\big)-\Pi\big(1+\frac{1}{\mu^2},\arcsin \big(\mu \sqrt{\frac{\coth^2\frac{\tau}{2}}{1+\mu^2}}\big),1-\frac{1}{\mu^4}\big)\bigg)\bigg)\cdot\nonumber \\&[\Theta(\tau-\mathrm{arccosh}(2\mu^2-1))-\Theta(\tau-\mathrm{arccosh(2\mu^2+1)})]\nonumber \\&+(1-2\mu^2\frac{2\cosh\tau}{\cosh2\tau-1})\Theta(\tau-\mathrm{arccosh}(1+2\mu^2))\bigg]\,.
\end{eqnarray}
}
The expression for $\mu < 1$ has a similar form; we omit it for space reasons. 
The results can be integrated numerically for any value of $\mu$.
We can immediately plot the pairs $(\mu,e^{\phi_{IR}})$, see figure \ref{fig: dynamics_linear_quiver_1}.
Nevertheless, to make a qualitative study of the behavior of $e^{\phi_{IR}}$ one can replace the flavor distribution by one described by a Heaviside theta function. After all, the previous distribution can be interpreted simply as a smoothed out Heaviside theta function. This consideration will be addressed in the next section.

As discussed in \cite{Bigazzi:2008qq}, keeping constant the value of $h_0$ amounts to fix the glueball and KK scale. In this case, and being $e^{\phi_{IR}}$ inversely proportional to $N_f$, we immediately see that the Hagedorn temperature decreases for increasing number of flavors, in agreement with the behavior displayed by the WSS and MN models addressed above.

\subsubsection{Qualitative behavior of $T_{H}$ in the Klebanov-Strassler theory}
Consider the flavor distribution given by
\begin{eqnarray}
    N_f(\tau)=N_f\Theta(\tau_0-\tau_q)\,,
\end{eqnarray}
where $\Theta(\tau)$ is the Heaviside theta function. Then
\begin{eqnarray}
e^{-\phi_{IR}}=\frac{1}{4\pi}\int_{\tau_q}^{\tau_0}N_f \Theta (\tau_0-\tau_q)=\frac{N_f}{4\pi}(\tau_0-\tau_q)\,.
\end{eqnarray}
Let us parameterize $\tau_0$ as $\tau_0=\frac{4\pi}{N_f}+\tau_a$, where $\tau_a$ does not depend on $N_f$, and can be interpreted as a UV scale at which we impose the matching with the unflavored theory. Then we have
\begin{eqnarray}
    e^{-\phi_{IR}}=1+\frac{N_f}{4\pi}(\tau_a-\tau_q)\,.
\end{eqnarray}
This function goes to one for both $N_f \rightarrow 0$ and for large masses $\tau_q\rightarrow \tau_a$ (the matching point). Then, the next to leading order correction to $T_H$ goes as $\sim N_f(\tau_a-\tau_q)$ which decreases as $N_f$ increases and the quark mass decreases.

To get an even more explicit realization of the latter statement in terms of field theory quantities, let us consider the case where both $\tau_a$ and $\tau_q$ are large. In this regime, using the holographic radius/energy relation $e^{\tau/3}\approx \mu/\Lambda$ (where $\Lambda$ is the IR scale) we get
\be
e^{-\phi_{IR}}\approx 1+\frac{3 N_f}{4\pi}\log\frac{\Lambda_a}{m_q}\,,
\ee
where $\Lambda_a$ is a UV scale and $m_q<\Lambda_a$ is the flavor mass.  Hence, to leading order in the flavor backreaction, and plugging back $g_s$ such that $M \rightarrow g_s M = \lambda$ (the 't Hooft coupling) and $N_f \rightarrow g_s N_f = \lambda N_f / M$,
\begin{eqnarray}
\frac{T_H}{\sqrt{T_s}}\approx \frac{1}{2\sqrt{\pi}}+\frac{\sqrt{3}}{\sqrt{2^5\pi \lambda} h_0^\frac{3}{4}}\left[1-\frac{3}{16\pi} \lambda \frac{N_f}{M}\log\frac{\Lambda_a}{m_q}\right]\,.
\end{eqnarray}

\section{Conclusions}

In the first part of this paper we have proposed an eleven-dimensional effective action to compute the Hagedorn temperature $T_H$ in confining theories with a dual gravity description.
Working in the M-theory setting allows to extend the known computations to a wider regime of parameters. 
In particular, the novel results concern the large $g_s$, or equivalently the large $\eta \sim \lambda^a/N$ regime.\footnote{The value of $a>0$ depends on the specific theory under scrutiny.}
The results have been derived at leading and NL orders in the expansion in $1/\lambda$.\footnote{At NNLO the results of this paper are expected to be incomplete.}
We have presented our general formulae in the ABJM and WYM theories.

The recent interest in studying the Hagedorn temperature in holography has been boosted by the possibility of comparing the stringy results with predictions from Quantum Spectral Curve methods.
The latter has been applied to ${\cal N}=4$ SYM and ABJM \cite{Harmark:2017yrv, Harmark:2018red, Harmark:2021qma, Ekhammar:2023glu, Ekhammar:2023cuj}, but only in the small $\eta$ limit.
It would be clearly interesting to extend these methods to the large $\eta$ regime, or to the first non-perturbative corrections in the small $\eta$ regime, and compare the results with our predictions for the ABJM case.
Other directions to pursue are to apply our results to the Type IIB cases and supplement our effective results with the ones coming from the membrane quantization, which are expected to become essential from NNLO on.

In the WYM example we have found that the ratio $T_H/T_c$ for ``realistic'' values of the coupling is close to the one expected in pure YM from the lattice.
We have also studied the behavior of $T_H/\sqrt{T_s}$ with the $\theta$-angle, finding that it increases with $\theta$ at fixed coupling.

In the second part of the paper we have calculated the effect of dynamical flavors on the value of $T_H$ in holographic settings.
The existing models allowing for this calculation are few, but the results are qualitatively homogeneous.
The value of $T_H$ normalized with the string tension is found to be affected by the flavors starting at NL order, and the effect is always a reduction of $T_H$ at fixed coupling.
The magnitude of the reduction depends on the number and mass of the flavors, as expected.

It would be interesting to check these behaviors with other methods, for example on the lattice.

\vskip 15pt \centerline{\bf Acknowledgments} \vskip 10pt 

\noindent 
We thank Michele Caselle for raising our interest in the flavor contribution to the Hagedorn temperature and Jorge Russo and Wolfgang M\"uck for very useful suggestions.


\appendix

\section{Flavor effects on CFTs on spheres}
\label{sec:appendixCFTs} 

In this appendix we consider the flavor effects on $T_H$ in some CFTs compactified on spheres, to highlight the different behavior with respect to the one encountered in truly confining models in the main body of the paper.

\subsection{Flavored ABJM model}

In this case the flavored smeared solution in string frame is contained in \cite{Conde:2011sw} (section 6).
Its form is very simple, since it is still an AdS solution with constant dilaton, RR two-form and four-form.
For what concerns the calculation of $T_H$ the flavors enter only the value of the parameter $l$ of formula (\ref{muhere}), which for the AdS cases is just the AdS radius.
This is called $L$ in \cite{Conde:2011sw}.
Solving (\ref{NLO}) at NLO we find
\begin{equation}
	\frac{T_H}{\sqrt{T_s}} = \frac{1}{2\sqrt{\pi}} + \frac{3}{4\sqrt{2\pi}L} \,.
\end{equation}
The dependence of $L$ on the flavors is derived through the definitions
\begin{equation}
	L^4 = 2\pi^2 \frac{N}{k} \frac{(2-q)b^4}{q(q+\eta q-\eta)}\,, \qquad b=\frac{q(q+\eta)}{2(q+\eta q-\eta)}\,, \qquad q=\frac{3(1+\eta)-\sqrt{9\eta^2-2\eta+9}}{2}\,,
\end{equation}
where 
\begin{equation}
	\eta=1+\frac34 \frac{N_f}{k}\,,
\end{equation}
$k$ is the CS level and $L^4 = 2\pi^2 N/k$ in the unflavored theory.

Now, while $l$ in the WSS, flavored MN and KS theories is an \emph{increasing} function of the number of flavors, such that the latter decrease $T_H/\sqrt{T_s}$, in the ABJM case $L$ is a \emph{decreasing} function of $N_f/k$.
Thus, the two classes of theories have opposite behaviors when flavors are added.

\subsection{Flavored Klebanov-Witten theory}

Another example comes from the flavored versions of the KW theory.
Let us consider the massive-flavor cases (the massless one is singular in the IR), for example the solution in \cite{Bigazzi:2008ie}.
In the IR it is still an AdS solution, with a rescaled radius with respect to the unflavored theory.
In string frame, the relevant part of the metric in \cite{Bigazzi:2008ie} is 
\begin{equation}
	ds^2 = e^{\phi/2} h^{-1/2} \left(-dt^2 +dx_idx^i \right)+  e^{\phi/2} h^{1/2}e^{2f}d\rho^2 + ... \,, 
\end{equation}
where the values of the functions are
\begin{eqnarray}
	&& h(\rho) = 27 \pi g_s N \alpha'^2 \frac{1}{a^4} \left(\int_{\rho_q}^0 e^{-4\rho_*}d\rho_* +\frac{e^{-4\rho}-e^{-4\rho_q}}{4}\right)\,, \qquad a= \frac{(2e^{3\rho_q}-2-6\rho_q)^{1/6}}{(1-e^{3\rho_q})^{1/3}}\,, \nonumber \\
	&& e^{\phi} = -\frac{4\pi}{g_s N_f (3\rho_q - e^{3\rho_q} +1)}\qquad ({\rm for}\ \rho < \rho_q)\,, \qquad \qquad e^{2f}= a^2 e^{2\rho}\,.
\end{eqnarray}
In this formula $\rho_q$ is a parameter, the radial position of the tip of the flavor branes, which carries the information about the mass of the flavors.
Performing the integral, in the IR ($\rho \rightarrow - \infty$) the function $h$ can be written as $h=h_0/a^4 e^{4\rho}$, where $h_0$ is a flavor-independent constant.
The metric reduces to
\begin{equation}
	ds^2 = e^{\phi/2} \left(\frac{h_0}{a^4} \right)^{-1/2} e^{2\rho} \left(-dt^2 +dx_idx^i \right)+  e^{\phi/2} h_0^{1/2}d\rho^2 + ... \,.
\end{equation}
Passing to the variable $r=a e^{\phi/2+\rho}$ we get the AdS metric
\begin{equation}
	ds^2 = \frac{r^2}{R^2}\left(-dt^2 +dx_idx^i \right) + \frac{R^2}{r^2} dr^2 + ... \,,
\end{equation}
with $R = R_{(N_f=0)} e^{\phi/4}$.

The behavior of the dilaton as a function of the flavors is not transparent. In fact, the parameter $\rho_q$ depends on the mass but also on $N_f$.\footnote{The mass $m_q$ of the quarks is given by the energy of a string stretched along the radial direction from $\rho_q$ to $\rho=-\infty$, giving the relation
	\begin{equation}
		m_q \sim \int_{-\infty}^{\rho_q}e^{\phi/2+f}d\rho = e^{\phi/2}a e^{\rho_q} \,.
	\end{equation}
}

A possible argument is as follows.
If we require that in the limit $g_sN_f \rightarrow 0$ we have $e^{\phi} \rightarrow 1$, we get $\rho_q \sim -4\pi/3 g_s N_f$.
Using the latter relation we get that $e^{\phi} > 1$ for $g_s N_f >0$. 
In this way, the flavors increase the AdS radius, decreasing $T_H$.
One can check that the same happens in the other flavored KW solution considered in \cite{Bigazzi:2008zt}.



\begin{thebibliography}{99}
	
\bibitem{Hagedorn:1965st}
R.~Hagedorn,
``Statistical thermodynamics of strong interactions at high-energies,''
Nuovo Cim. Suppl. \textbf{3} (1965), 147-186
CERN-TH-520.	

\bibitem{Huang:1970iq}
K.~Huang and S.~Weinberg,
``Ultimate temperature and the early universe,''
Phys. Rev. Lett. \textbf{25} (1970), 895-897.

\bibitem{Sundborg:1984uk}
B.~Sundborg,
``Thermodynamics of Superstrings at High-energy Densities,''
Nucl. Phys. B \textbf{254} (1985), 583-592.
	
\bibitem{Canneti:2024iyn}
T.~Canneti,
``On the asymptotic density of states in solvable models of strings,''
[arXiv:2406.08405 [hep-th]].	

\bibitem{Harmark:2021qma}
T.~Harmark and M.~Wilhelm,
``Solving the Hagedorn temperature of AdS$_{5}$/CFT$_{4}$ via the Quantum Spectral Curve: chemical potentials and deformations,''
JHEP \textbf{07} (2022), 136
[arXiv:2109.09761 [hep-th]].	

\bibitem{malda}J.~M.~Maldacena, unpublished.
	
\bibitem{Urbach:2022xzw}
E.~Y.~Urbach,
``String stars in anti de Sitter space,''
JHEP \textbf{04} (2022), 072
[arXiv:2202.06966 [hep-th]].	
		
\bibitem{Bigazzi:2022gal}
F.~Bigazzi, T.~Canneti and A.~L.~Cotrone,
``On the Hagedorn temperature in holographic confining gauge theories,''
JHEP \textbf{01} (2023), 034
[arXiv:2210.09893 [hep-th]].			

\bibitem{Urbach:2023npi}
E.~Y.~Urbach,
``The black hole/string transition in AdS$_{3}$ and confining backgrounds,''
JHEP \textbf{09} (2023), 156
[arXiv:2303.09567 [hep-th]].
		
\bibitem{Bigazzi:2023oqm}
F.~Bigazzi, T.~Canneti and W.~M\"uck,
``Semiclassical quantization of the superstring and Hagedorn temperature,''
JHEP \textbf{08} (2023), 185
[arXiv:2306.00588 [hep-th]].	
	
\bibitem{Ekhammar:2023glu}
S.~Ekhammar, J.~A.~Minahan and C.~Thull,
``The asymptotic form of the Hagedorn temperature in planar super Yang-Mills,''
J. Phys. A \textbf{56} (2023) no.43, 435401
[arXiv:2306.09883 [hep-th]].	

\bibitem{Bigazzi:2023hxt}
F.~Bigazzi, T.~Canneti and A.~L.~Cotrone,
``Higher Order Corrections to the Hagedorn Temperature at Strong Coupling,'' 
JHEP \textbf{10} (2023), 056
[arXiv:2306.17126 [hep-th]].

\bibitem{Ekhammar:2023cuj}
S.~Ekhammar, J.~A.~Minahan and C.~Thull,
``The ABJM Hagedorn Temperature from Integrability,''
JHEP \textbf{10} (2023), 066
[arXiv:2307.02350 [hep-th]].

\bibitem{Harmark:2024ioq}
T.~Harmark,
``Hagedorn temperature from the thermal scalar in AdS and pp-wave backgrounds,''
JHEP \textbf{06}, 140 (2024)
[arXiv:2402.06001 [hep-th]].

\bibitem{Bigazzi:2024biz}
F.~Bigazzi, T.~Canneti, F.~Castellani, A.~L.~Cotrone and W.~M\"uck,
``Hagedorn temperature in holography: world-sheet and effective approaches,''
[arXiv:2407.00375 [hep-th]].

\bibitem{Horowitz:1997jc}
G.~T.~Horowitz and J.~Polchinski,
``Selfgravitating fundamental strings,''
Phys. Rev. D \textbf{57} (1998), 2557-2563
[arXiv:hep-th/9707170 [hep-th]].

\bibitem{Atick:1988si}
J.~J.~Atick and E.~Witten,
``The Hagedorn Transition and the Number of Degrees of Freedom of String Theory,''
Nucl. Phys. B \textbf{310} (1988), 291-334.


\bibitem{Russo:2001vh}
J.~G.~Russo,
``Free energy and critical temperature in eleven-dimensions,''
Nucl. Phys. B \textbf{602} (2001), 109-131
[arXiv:hep-th/0101132 [hep-th]].

\bibitem{Aharony:2008ug}
O.~Aharony, O.~Bergman, D.~L.~Jafferis and J.~Maldacena,
``N=6 superconformal Chern-Simons-matter theories, M2-branes and their gravity duals,''
JHEP \textbf{10} (2008), 091
[arXiv:0806.1218 [hep-th]].

\bibitem{Witten:1998zw}
E.~Witten,
``Anti-de Sitter space, thermal phase transition, and confinement in gauge theories,''
Adv. Theor. Math. Phys. \textbf{2} (1998), 505-532
[arXiv:hep-th/9803131 [hep-th]].

\bibitem{Dubovsky:2011tu}
S.~Dubovsky, A.~Lawrence and M.~M.~Roberts,
``Axion monodromy in a model of holographic gluodynamics,''
JHEP \textbf{02} (2012), 053
[arXiv:1105.3740 [hep-th]].

\bibitem{Bigazzi:2015bna}
F.~Bigazzi, A.~L.~Cotrone and R.~Sisca,
``Notes on Theta Dependence in Holographic Yang-Mills,''
JHEP \textbf{08} (2015), 090
[arXiv:1506.03826 [hep-th]].

\bibitem{Bringoltz:2005xx}
B.~Bringoltz and M.~Teper,
``In search of a Hagedorn transition in SU(N) lattice gauge theories at large-N,''
Phys. Rev. D \textbf{73} (2006), 014517
[arXiv:hep-lat/0508021 [hep-lat]].

\bibitem{Arean:2010hu}
D.~Arean, E.~Conde, A.~V.~Ramallo and D.~Zoakos,
``Holographic duals of SQCD models in low dimensions,''
JHEP \textbf{06} (2010), 095.

\bibitem{Borsanyi:2013bia}
S.~Borsanyi, Z.~Fodor, C.~Hoelbling, S.~D.~Katz, S.~Krieg and K.~K.~Szabo,
``Full result for the QCD equation of state with 2+1 flavors,''
Phys. Lett. B \textbf{730} (2014), 99-104
[arXiv:1309.5258 [hep-lat]].

\bibitem{Caselle:2015tza}
M.~Caselle, A.~Nada and M.~Panero,
``Hagedorn spectrum and thermodynamics of SU(2) and SU(3) Yang-Mills theories,''
JHEP \textbf{07} (2015), 143
[erratum: JHEP \textbf{11} (2017), 016]
[arXiv:1505.01106 [hep-lat]].

\bibitem{Boyd:1996bx}
G.~Boyd, J.~Engels, F.~Karsch, E.~Laermann, C.~Legeland, M.~Lutgemeier and B.~Petersson,
``Thermodynamics of SU(3) lattice gauge theory,''
Nucl. Phys. B \textbf{469} (1996), 419-444
[arXiv:hep-lat/9602007 [hep-lat]].

\bibitem{Bigazzi:2014qsa}
F.~Bigazzi and A.~L.~Cotrone,
``Holographic QCD with Dynamical Flavors,''
JHEP \textbf{01} (2015), 104
[arXiv:1410.2443 [hep-th]].

\bibitem{Burrington:2007qd}
B.~A.~Burrington, V.~S.~Kaplunovsky and J.~Sonnenschein,
``Localized Backreacted Flavor Branes in Holographic QCD,''
JHEP \textbf{02} (2008), 001
[arXiv:0708.1234 [hep-th]].

\bibitem{Conde:2011rg}
E.~Conde, J.~Gaillard and A.~V.~Ramallo,
``On the holographic dual of $N=1$ SQCD with massive flavors,''
JHEP \textbf{10} (2011), 023
[erratum: JHEP \textbf{08}, 082 (2013)]
[arXiv:1107.3803 [hep-th]].

\bibitem{Klebanov:2000hb}
I.~R.~Klebanov and M.~J.~Strassler,
``Supergravity and a confining gauge theory: Duality cascades and chi SB resolution of naked singularities,''
JHEP \textbf{08} (2000), 052
[arXiv:hep-th/0007191 [hep-th]].

\bibitem{Bigazzi:2008qq}
F.~Bigazzi, A.~L.~Cotrone, A.~Paredes and A.~V.~Ramallo,
``The Klebanov-Strassler model with massive dynamical flavors,''
JHEP \textbf{03} (2009), 153
[arXiv:0812.3399 [hep-th]].

\bibitem{Harmark:2017yrv}
T.~Harmark and M.~Wilhelm,
``Hagedorn Temperature of AdS$_5$/CFT$_4$ via Integrability,''
Phys. Rev. Lett. \textbf{120} (2018) no.7, 071605
[arXiv:1706.03074 [hep-th]].

\bibitem{Harmark:2018red}
T.~Harmark and M.~Wilhelm,
``The Hagedorn temperature of AdS$_5$/CFT$_4$ at finite coupling via the Quantum Spectral Curve,''
Phys. Lett. B \textbf{786} (2018), 53-58
[arXiv:1803.04416 [hep-th]].

\bibitem{Conde:2011sw}
E.~Conde and A.~V.~Ramallo,
``On the gravity dual of Chern-Simons-matter theories with unquenched flavor,''
JHEP \textbf{07} (2011), 099
[arXiv:1105.6045 [hep-th]].

\bibitem{Bigazzi:2008ie}
F.~Bigazzi, A.~L.~Cotrone, A.~Paredes and A.~Ramallo,
``Non chiral dynamical flavors and screening on the conifold,''
Fortsch. Phys. \textbf{57} (2009), 514-520
[arXiv:0810.5220 [hep-th]].

\bibitem{Bigazzi:2008zt}
F.~Bigazzi, A.~L.~Cotrone and A.~Paredes,
``Klebanov-Witten theory with massive dynamical flavors,''
JHEP \textbf{09} (2008), 048
[arXiv:0807.0298 [hep-th]].


\end{thebibliography}
\end{document}